\documentclass[10pt]{article}
\pdfoutput=1
\usepackage{jheparxiv}
\usepackage[latin1]{inputenc}
\usepackage{latexsym,diagbox,stackrel}
\usepackage{relsize}
\usepackage{shuffle}
\usepackage{amsmath}
\usepackage{mathrsfs}
\usepackage{dsfont}
\usepackage{array}
\usepackage{nicefrac}
\usepackage{arydshln,leftidx,mathtools}
\usepackage{bbold}
\usepackage{graphicx}
\usepackage{subcaption}
\usepackage{pdflscape}
\usepackage{amssymb}
\usepackage{multirow}
\usepackage{enumerate}
\usepackage{blkarray}
\usepackage[dvipsnames,svgnames,table]{xcolor}
\usepackage{cleveref}
\usepackage{tikz}
\usepackage{graphicx}
\usepackage{slashed}
\usetikzlibrary{shapes.geometric, arrows}
\definecolor{light-gray}{gray}{0.95}
\definecolor{light-grayII}{gray}{0.85}
\usepackage{array}
\usepackage{pdfpages}
\usepackage{empheq}
\usepackage{comment}
\usepackage{graphbox}

\def\d{\text{d}}

\newcommand{\cI}{\mathcal{I}}

\newcommand{\SL}{\mathrm{SL}}

\newcommand{\tr}{\mathrm{tr}}

\newcommand{\vol}{\mathrm{vol}\,}

\newcommand{\KN}{\text{KN}}

\newcommand{\eg}[2]{g^{(#1)}_{#2}}
\newcommand{\egg}[1]{g^{(1)}_{#1}}
\newcommand{\reg}{\textbf{Reg}}

\renewcommand{\[}{\begin{equation}\begin{aligned}}
\renewcommand{\]}{\end{aligned}\end{equation}}

\subheader{\hfill \begin{tabular}{r} \texttt{QMUL-PH-25-24} \end{tabular}}
 
\title{\Large Superstring amplitudes from the field-theory limit: an $n$-point map at one loop}
\author{Ricardo Monteiro}
\author{\& Lecheng Ren}

\affiliation{Centre for Theoretical Physics, Department of Physics and Astronomy, \\
        Queen Mary University of London, E1 4NS, United Kingdom}
        
\emailAdd{ricardo.monteiro@qmul.ac.uk}
\emailAdd{lecheng.ren@qmul.ac.uk}

\abstract{
Are perturbative superstring amplitudes for massless external states just an $\alpha'$ dressing of super-Yang-Mills/supergravity? This is the case at tree level, where the worldsheet correlators at $n$ points can be written in a natural worldsheet basis, such that the kinematic coefficients are BCJ numerators of super-Yang-Mills/supergravity amplitudes, with the non-trivial $\alpha'$ dependence carried only by the Koba-Nielsen factor. Motivated by this construction, we present for the first time a complete worldsheet basis of one-loop superstring correlators at $n$ points. All the kinematic coefficients associated to non-cusp basis elements are identified with pieces of one-loop BCJ numerators of super-Yang-Mills/supergravity. This determines the superstring correlators up to 15 points in terms of field theory. Starting at 16 points (modular weight 12), the worldsheet basis may include cusp forms, which vanish in the field-theory degeneration, such that the associated coefficients cannot be fixed in this manner. Therefore, the one-loop answer to our initial question is determined, at high multiplicity, by whether the coefficients of cusp basis elements vanish or not. As a by-product of our construction, we present new constraints on the field-theory limit that result from string modularity. These are expressed as additional relations among one-loop BCJ numerators in maximal super-Yang-Mills/supergravity starting at 6 points.
}

\begin{document}

\maketitle


\section{Introduction}

The limit $\alpha'\to0$ of string-theory amplitudes has long been a useful tool to construct field-theory amplitudes, e.g.~\cite{Green:1982sw,Bern:1991aq}. At first glance, it would be unexpected if the reverse path could be followed: given its parameter $\alpha'$, string-theory amplitudes appear to require more information for their construction than field-theory amplitudes. And yet, if we focus on the moduli-space {\it integrand} of superstring amplitudes, the structure of the $\alpha'$ dependence of the worldsheet correlation functions for the scattering of massless external states is very simple, at least at low loop order: the non-trivial dependence on $\alpha'$ is carried solely by a known Koba-Nielsen factor (the trivial dependence on $\alpha'$ being the overall normalisation of the amplitude, set by dimensional analysis). The suggestion is that the superstring correlator is potentially constructible from the field-theory limit $\alpha'\to0$. While perhaps not widely appreciated, this is known to be the case at tree level: for instance, the closed superstring amplitude for massless external states can be written as \cite{Mafra:2011nv,Mafra:2011kj}
\[
\label{eq:Atree}
\mathcal{A}_n^{(0),\text{closed}} = \alpha'^{n-2}\int_{\mathcal{M}_{0,n}} \frac{d^{2n}\sigma}{\vol \SL(2,\mathbb C)}\; \Bigg| \sum_{\rho\in S_{n-2}}\frac{N(1,\rho(2),\cdots,\rho(n-1),n)}{\sigma_{1\rho(2)}\sigma_{\rho(2)\rho(3)}\cdots\sigma_{\rho(n-1)n}\sigma_{n1}}\Bigg|^2\;
\prod_{i<j}|\sigma_{ij}|^{\alpha' p_i\cdot p_j} \,,
\]
where $\sigma_{ij}=\sigma_i-\sigma_j$. The numerators $N$ are functions of the kinematics of the external states (momenta and polarisations), and we allow for distinct polarisations in the second copy of the conjugation, e.g.~$ \,\overline{N(\{p_i,\epsilon_i\})} \mapsto N(\{p_i,\tilde \epsilon_i\})\,$ so that the scattered states have polarisation $\epsilon_i \tilde \epsilon_i$. The point is that these numerators can be extracted from supergravity amplitudes with the same external states. To compute the string amplitude, there is still the very challenging step of performing the moduli-space integral. Nevertheless, we conclude that the field-theory information, inserted in a suitable worldsheet basis and dressed by the Koba-Nielsen factor, is sufficient to determine the superstring correlator.

This paper is part of a program \cite{Geyer:2021oox,Geyer:2024oeu} to study superstring amplitudes at loop level (for now, only their moduli-space integrands) by exploiting the field-theory limit. By the latter, we really mean the field-theory loop integrand, given that the loop integration leads in 10D field theory to ultraviolet divergence.

The first step in this program at a given loop order is to find a general representation of the moduli-space integral, e.g.~the chiral-splitting representation at low loop order \cite{DHoker:1988pdl,DHoker:1989cxq}. The second step is to find a natural basis for the worldsheet dependence of the superstring correlator, i.e.~the analogue of the tree-level `Parke-Taylor denominators' in \eqref{eq:Atree}. And the third step is to identify the kinematic coefficients in that basis with field-theory (integrand-level) objects, i.e.~the analogue of the numerators $N$ in \eqref{eq:Atree}. Ref.~\cite{Geyer:2021oox} showed how to follow these steps at four points up to three loops. The three-loop result provided a novel conjecture (given that the completeness of the basis was conjectural), which is consistent with a previously known piece contributing at low energies, derived from the pure-spinor formalism~\cite{Gomez:2013sla}. Continuing this program, ref.~\cite{Geyer:2024oeu} focused on one-loop amplitudes at higher multiplicity, with general results up to seven points, and some results at higher points. In this paper, we will generalise these one-loop results, aided by a new understanding of the $n$-point worldsheet basis.

The most convenient field-theory objects for constructing string-theory amplitudes appear to be BCJ numerators, which in field theory are a choice of numerators of Feynman-like diagrams with particular algebraic properties \cite{Bern:2008qj,Bern:2010ue}. The numerators $N$ in \eqref{eq:Atree} are precisely such objects, at tree level. Introduced by Bern, Carrasco and Johansson, the BCJ numerators are famous for relating gauge-theory amplitudes to gravity ones via a `double copy', with applications ranging from ultraviolet studies of supergravity to new methods in classical general relativity; see e.g.~\cite{Bern:2019prr,Kosower:2022yvp,Adamo:2022dcm}. The historical origin of the `double copy' lies in string theory, where it is realised by the chiral splitting between left- and right-movers, as first exploited for tree-level amplitudes in the KLT relations~\cite{Kawai:1985xq}.
At loop level, the BCJ numerators are integrand-level objects, and while there are many results, there is no known algorithm to derive loop-level numerators to arbitrary multiplicity, with a few exceptions at one loop for the simplest helicity configurations in 4D \cite{Boels:2013bi,Monteiro:2011pc,He:2015wgf}. In fact, there are apparent obstructions at higher loops, depending also on multiplicity and degree of supersymmetry, which led to some workarounds being devised \cite{Bern:2017yxu,Bern:2018jmv,Bern:2024vqs,Mogull:2015adi}. Chiral splitting in superstring theory \cite{DHoker:1988pdl,DHoker:1989cxq} is also not straightforward at higher genus \cite{Donagi:2013dua}. It is certainly of interest to understand when we can expect the existence of BCJ numerators, and what is the optimal replacement when there are obstructions. Fortunately for this paper, as argued in \cite{Geyer:2024oeu} and complemented here, their existence at one loop is guaranteed by the general form of the superstring amplitude, due to the validity of chiral splitting at genus one, and to the regularity of the field-theory limit at integrand level. This does not mean, however, that the explicit construction of one-loop BCJ numerators for super-Yang-Mills/supergravity is straightforward: the state-of-the-art is seven points \cite{Edison:2022jln}. So we will not be presenting complete expressions for superstring correlators. Rather, we will show how the kinematic coefficients of the correlators in our worldsheet basis can be identified with BCJ numerators (up to the issue of cusp forms).

To emphasise the two-ways connection between string and particle amplitudes, we will not only learn about the former from the latter, which was our original motivation, but will also learn about the latter from the former. In particular, we will identify new constraints that the BCJ numerators in maximal super-Yang-Mills/supergravity must satisfy starting at 6 points, which follow from the modularity of the superstring correlator. At present, the physical interpretation of these constraints in field theory is unclear to us.

Before proceeding, we want to mention two lines of work that provided crucial clues to our results. The first is the longstanding study of one-loop superstring amplitudes by Mafra, Schlotterer and collaborators \cite{Mafra:2012kh,Mafra:2014oia,Mafra:2014gja,Broedel:2014vla,Mafra:2017ioj,Mafra:2018nla,Mafra:2018pll,Mafra:2018qqe}, mostly based on explicit computations with the pure-spinor formalism \cite{Berkovits:2000fe,Berkovits:2002zk}. Ref.~\cite{Mafra:2017ioj} is particularly relevant, as we will see. The second line of work is the string-like formulation of field theory provided by the ambitwistor string \cite{Mason:2013sva,Adamo:2013tsa,Casali:2014hfa,Geyer:2015bja,Geyer:2015jch,Geyer:2016wjx,Geyer:2017ela,Geyer:2018xwu,Geyer:2022cey} and its predecessors \cite{Witten:2003nn,Roiban:2004yf,Cachazo:2013hca,Cachazo:2013iea}; see also \cite{Skinner:2013xp,Berkovits:2013xba,Adamo:2015hoa,He:2015yua,Cachazo:2015aol,Feng:2016nrf,Cardona:2016bpi,Cardona:2016gon,Chen:2016fgi,Gomez:2016cqb,Gomez:2017lhy,He:2017spx,Gomez:2017cpe,Roehrig:2017gbt,Ahmadiniaz:2018nvr,Edison:2020uzf,Abhishek:2020sdr,Kalyanapuram:2021xow,Kalyanapuram:2021vjt,Feng:2022wee,Dong:2023stt,Xie:2024pro} for relevant work. The picture introduced in \cite{Geyer:2015bja} of the field-theory worldsheet as a nodal Riemann sphere, with the node representing the loop momentum, was a major inspiration for our program.

This paper is organised as follows. In section~\ref{sec:review}, we review the chiral-splitting formalism for one-loop superstring amplitudes, the structure of the 
chiral worldsheet correlator, and its connection to the BCJ numerators in field theory. In section~\ref{sec:elliptic}, we describe the basic features of the worldsheet basis we employ to construct the correlator. In the sections~\ref{sec:4pt} to \ref{sec:8pt}, we cover the cases of 4-to-8-point correlators. In section~\ref{sec:npt}, we give a complete description of the worldsheet basis at $n$ points, describing how to fix the correlators from the field-theory limit, up to an ambiguity caused by cusp forms starting at 16 points. In section~\ref{sec:G2}, we discuss the general constraints imposed on field theory by the modularity of the superstring correlator, building on earlier sections starting at 6 points. We conclude with a discussion of our results in section~\ref{sec:conclusion}. Finally, we include some technical appendices.


\section{Review}
\label{sec:review}

In this section, we briefly review the structure of one-loop superstring amplitudes and their field-theory limit, and in particular the approach of ref.~\cite{Geyer:2024oeu}, which will be improved on and generalised in this paper.

\subsection{Structure of the superstring amplitude}

We employ the chirally-split description of the superstring amplitude  \cite{DHoker:1988pdl,DHoker:1989cxq}. The idea is to maintain the chiral/anti-chiral factorisation of the moduli-space integrand, prior to the integration over the zero-mode of $\partial X^\mu$; this zero mode is identified with the loop momentum in the field-theory limit. We have at one loop, for the closed string,
\begin{equation}
\label{eq:Ac}
 \mathcal{A}_n^{(1),\text{closed}} = \alpha'^n \int_{\mathbb R^D} d^D\ell \int_{\mathcal{F}} d^2\tau \int_{T^{n-1}}\!d^2z_2\cdots  d^2z_n\;\;\mathcal{I}_n(\ell)\;\tilde{\mathcal{I}}_n(\ell)\;\big|\text{KN}_n(\ell)\big|^2 \,,
\end{equation}
while for the open string,
\begin{equation}
\label{eq:Ao}
 \mathcal{A}_n^{(1),\text{open}} = \alpha'^n \sum_\text{top} C_\text{top} \int_{\mathbb R^D} d^D\ell\int_{\mathcal{D}_\text{top}} d\tau \int_{O^{n-1}_\text{top}}\!d^2z_2\cdots  d^2z_n\;\;\mathcal{I}_n(\ell)\;\big|\text{KN}^{(\alpha'\mapsto4\,\alpha')}_n(\ell)\big| \,.
\end{equation}
The integration domains are the standard ones. In the open string case, we have a sum over topologies (cylinder and Moebius-strip) with corresponding Chan-Paton colour factors $C_\text{top}$. The chiral Koba-Nielsen factor is given by
\begin{equation}
\label{eq:KN}
 \text{KN}_n(\ell) = \exp\frac{\alpha'}{2}\Bigg(\sum_{1\leq i<j\leq n} p_i\cdot p_j\ln \theta_1(z_{ij},\tau) +2i\pi\,\ell\cdot\sum_{j=1}^n z_j\, p_j \,+i\pi\tau\,\ell^2\Bigg)\,,
\end{equation}
where the odd Jacobi theta function reads
\[
\theta_1(z,\tau) := 2q^{1/8} \sin(\pi z) \prod^\infty_{n=1} (1-q^n) (1-q^n e^{2\pi i z})(1-q^n e^{-2\pi iz})\,,
 \quad \text{with} \quad q:=e^{2\pi i\tau}\,.
\]
Our goal is to construct the object $\mathcal{I}_n(\ell)$ --- which we refer to as the chiral integrand --- based on information from the field-theory limit. Its doubled appearance in the closed string is a realisation of the double copy.\footnote{As usual, the straightforward factorisation between the chiral $\mathcal{I}_n$ and the anti-chiral $\tilde{\mathcal{I}}_n$ in \eqref{eq:Ac} assumes a basis of external states that are factorised, e.g.~$\varepsilon_i^{\mu\nu}=\epsilon_i^\mu\tilde\epsilon_i^\nu$, with $\mathcal{I}_n$ dependent on $\epsilon_i$ and $\tilde{\mathcal{I}}_n$ dependent on $\tilde\epsilon_i$. Notice the analogy with the tree-level expression \eqref{eq:Atree}.}

One crucial property of the superstring amplitude is that there is a valid choice of $\mathcal{I}_n(\ell)$ that is independent of $\alpha'$ if the external states are massless, such that the dependence on $\alpha'$ of the amplitude is entirely given by the Koba-Nielsen factor and by the overall normalisation that we wrote explicitly. This is implied by the detailed form of the massless vertex operators and the OPE contractions leading to the worldsheet correlator. Close inspection indicates that, after factoring out the overall $\alpha'$ normalisation and the Koba-Nielsen factor, we are left with $\mathcal{I}_n(\ell)$ being a polynomial in $1/\alpha'$. The regularity of the field-theory limit then implies that any inverse power of $\alpha'$ can be resolved away, up to integration-by-parts in moduli space. The basis that we will study on the worldsheet is tied up with the choice of $\mathcal{I}_n(\ell)$ being independent of $\alpha'$. We leave more comments on the $\alpha'$ dependence to the appendix~\ref{app:Ialpha}, and we shall also return to this point in section~\ref{sec:elliptic}.

The procedure followed in ref.~\cite{Geyer:2024oeu} to determine $\mathcal{I}_n(\ell)$ starts with an ansatz: we write down the most general polynomial of a set of building blocks, consistent with worldsheet modularity, and then fix the kinematic coefficients of this polynomial using the field-theory limit. The building blocks, and their weights associated to modularity, are the following \cite{Broedel:2014vla,Mafra:2017ioj}.
\[
\label{eq:tablew}
\begin{tabular}{|c|c|c|c|}
\hline
object & weight \\
\hline
$2\pi i\,\ell_\mu$ & 1 \\
\hline
$ g^{(w)}_{ij}$ & $w$ \\
\hline
$G_{2K}$ & $2K$ \\
\hline
\end{tabular}
\]
We have already encountered the loop momentum $\ell_\mu$, so let us describe the other building blocks. We use the shorthand notation
\[
 g^{(w)}_{ij} = g^{(w)}(z_i-z_j,\tau) \,,
 \]
where the functions $g^{(w)}$ are defined from the Kronecker-Eisenstein series,
\begin{equation}
\label{eq:F}
 F(z,\eta, \tau):=  \frac{\theta_1'(0,\tau)\,\theta_1(z+\eta,\tau)}{\theta_1(\eta,\tau)\, \theta_1(z,\tau)} = \sum_{w=0}^\infty \eta^{w-1}g^{(w)}(z,\tau)\,.
\end{equation}
We have $\,g^{(0)}(z,\tau)= 1\,$,
\begin{equation}
\label{eq:g1g2}
  g^{(1)}(z,\tau) = \partial_z\ln\theta_1(z,\tau)\,,\quad\; g^{(2)}(z,\tau) = \frac1{2} \left( \big(\partial_z\ln\theta_1(z,\tau)\big)^2 +\partial_z^2 \ln\theta_1(z,\tau)\right)-\frac{\partial_z^3\theta_1(0,\tau)}{3!\,\partial_z\theta_1(0,\tau)}\,,
\end{equation}
and so on. Note that $g^{(w)}_{ij} =(-1)^w g^{(w)}_{ji}$. Only $g^{(1)}$ has a pole at $z=0$, with unit residue. There are relations among the $g^{(w)}_{ij}$, known as Fay relations, which result from the expansion in small $\eta_1,\eta_2$ of the identity
\[
\label{eq:Fay}
F(z,\eta_1,\tau)F(z',\eta_2,\tau) = F(z,\eta_1+\eta_2,\tau)F(z'-z,\eta_2,\tau)+F(z',\eta_1+\eta_2,\tau)F(z-z',\eta_1,\tau)\,.
\]
In the table \eqref{eq:tablew}, the functions $g^{(w)}_{ij}$ are attributed weight $w$, but they do not have a well-defined modular weight. In fact, they are not doubly-periodic,\footnote{Note that: $F(z+1,\eta,\tau) = F(z,\eta,\tau)\,,\; F(z+\tau,\eta,\tau)= e^{-2i\pi\eta}F(z,\eta,\tau)\,.$ So $F(z,\eta,\tau)$ is quasi-doubly-periodic.} as
\[
g^{(w)}(z+\tau,\tau) = \sum_{m=0}^w \frac{(-2\pi i)^m}{m!}\, g^{(w-m)}(z,\tau)\,.
\]
Modularity of the worldsheet correlator requires that
\[
\text{weight}(\mathcal{I}_n) = n-4\,,
\]
and that $\mathcal{I}_n(\ell)$ is invariant under the transformation
\[
\label{eq:monodromy}
i\text{-particle monodromy:} \qquad \ell_\mu \mapsto \ell_\mu - p_{i,\mu}\,, \qquad 
z_i\mapsto z_i+\tau\,.
\]
Regarding the first condition, on the weight of $\mathcal{I}_n$, it implies, for instance, that the chiral integrand at 4 points does not admit any of the building blocks in table \eqref{eq:tablew}, and therefore it must be a constant on the genus-one surface, i.e.~it is just a function of the momenta and polarisations of the external states.
Regarding the second condition, the monodromy transformations of the $g^{(w)}_{ij}$ and of the loop momentum must cancel overall in $\mathcal{I}_n$, which can be seen as a linear constraint among the coefficients of an ansatz. That was the approach taken in ref.~\cite{Geyer:2024oeu}. An alternative approach, which we will use here following \cite{Mafra:2017ioj}, is to employ a basis of worldsheet objects that are monodromy-invariant to start with, and where the Fay identities are already taken into account. Independently of how monodromy invariance is imposed, it can be shown that it leads to a modular worldsheet correlator after integration over $\ell$ \cite{Mafra:2018pll,Mafra:2018qqe}.

A related constraint on the chiral integrand is that $\mathcal{I}_n\,dz_2\cdots dz_n$ should have only logarithmic singularities. This excludes contributions to $\mathcal{I}_n$ such as $(g^{(1)}_{ij})^2$, which extends to the exclusion of `closed cycles' $g^{(1)}_{i_1i_2}g^{(1)}_{i_2i_3}\cdots g^{(1)}_{i_mi_1}$, which in turn extends to higher weights. The rule is:
\[
\textrm{exclude `closed cycles'} \quad  g^{(w_1)}_{i_1i_2}g^{(w_2)}_{i_2i_3}\cdots g^{(w_m)}_{i_mi_1}\,,
\]
with $m\geqslant 2$ and $w_a\geqslant1$. The exclusion can be understood in terms of Fay relations and integration-by-parts identities involving derivatives of the $g^{(w)}_{ij}$, though at present we do not have a general proof. We will mention examples at the end of section~\ref{sec:elliptic} (and in appendix~\ref{app:ibp7pt}).

Finally, we also included as building blocks in the table \eqref{eq:tablew} the holomorphic Eisenstein series,
\[
\label{eq:G2k}
G_{2K}(\tau):= \sum_{(m,n)\in\mathbb{Z}^2\setminus (0,0)} \frac{1}{(m+\tau n)^{2K}} \;=\; - g^{(2K)}(0,\tau)\,, \quad K\geq2\,.
\]
These objects are modular forms under SL(2,$\mathbb Z$) of weight $2K$, and can be written as polynomials (with rational coefficients) in the first two elements, $G_4$ and $G_6$.\footnote{Hence, we could have included only $G_4$ and $G_6$ in the table \eqref{eq:tablew}. As we will see later, however, it is convenient to consider the particular polynomials $G_{2K}$.} Since the chiral integrand $\mathcal{I}_n$ has weight $n-4$, these objects may appear first at 8 points, namely via $G_4$. They play an important role in this paper. We will explain for the first time how to constrain the corresponding contributions to the chiral integrand. There is an important remark, however, which puts into question the idea that the ansatz can be fully determined solely by the field-theory limit at high multiplicity. Cusp forms are polynomials of $G_4$ and $G_6$ with well-defined modular weight that vanish as $\tau\to i\infty$, which is the field-theory degeneration. So the field-theory limit is not sensitive to cusp forms. It is easy to see that this may occur first at 16 points, because $(G_4)^3$ and $(G_6)^2$ have both weight $12=16-4$; each has a finite degeneration limit, so that a linear combination of them vanishes in this limit. This point will be discussed in section~\ref{subsec:cusp}.

One curiosity that turns out to be important is the condition $K\geq2$ in \eqref{eq:G2k}. The would-be element
\[
\label{eq:G2}
G_{2}(\tau):= - g^{(2)}(0,\tau) = - \frac{\partial_z^3\theta_1(0,\tau)}{3\,\partial_z\theta_1(0,\tau)}
\]
is {\it not} a modular form.\footnote{While \,$G_{2K} \left( \frac{a\tau + b}{c\tau + d} \right) = (c\tau + d)^{2K} G_{2K}(\tau)$\, for $K\geq 2$, we have $G_2\left( \frac{a\tau + b}{c\tau + d} \right) = (c\tau + d)^2 G_2(\tau) - \frac{2\pi i c}{c\tau + d}$.} Interestingly, when we introduce $G_2$ into the superstring ansatz (forgetting for a moment that it should not be present), we can identify a would-be associated contribution to the field-theory limit. The absence of terms with $G_2$ in the actual superstring correlator implies that a piece of the field-theory loop integrand that we would naively expect from a purely field-theory perspective actually vanishes. From the would-be modular weight $2=6-4$, we can anticipate that this feature occurs first at 6 points. This non-trivial constraint on field theory, resulting from the modularity of the superstring, will be discussed in various examples, with the general $(n\geq6)$-point statement given in section~\ref{sec:G2}.

\subsection{The field-theory limit and BCJ numerators}
\label{subsec:FTlimit}

We want to construct the chiral integrand $\mathcal{I}_n$ using the building blocks in table \eqref{eq:tablew}, such that it has weight $n-4$ and is invariant under monodromy transformations \eqref{eq:monodromy}. 
After constructing such an ansatz, we wish to fix its coefficients using the field-theory limit. This corresponds to the degeneration $q:=e^{2\pi i\tau}\to0$; see e.g.~\cite{Green:1982sw,Tourkine:2013rda}. Thinking of the torus, we find it convenient to consider the Schottky parametrisation, by employing the coordinate transformation
\begin{equation}
e^{2i\pi z} = \frac{(\sigma-\sigma_+)(\sigma_*-\sigma_-)}{(\sigma-\sigma_-)(\sigma_*-\sigma_+)}
\,,\qquad 
dz = \left(\frac{1}{\sigma - \sigma_+}-\frac{1}{\sigma - \sigma_-}\right)\frac{d\sigma}{2\pi i}\,.
\end{equation}
The additional pair of marked points $\sigma_\pm$ is absent from the genus-one worldsheet.\footnote{Disks around $\sigma_+$ and $\sigma_-$ are excised from the Riemann sphere, and their boundaries are identified in such a manner that they form a handle. Notice also that the arbitrary choice of $\sigma_*$ drops out of both $\d z$ and $z_i-z_j$.} We note the identity
\[
(2\pi i)^{n-1}\prod_{i=2}^n dz_i = \prod_{i=2}^n \left(d\sigma_i \, \frac{\sigma_{+-}}{\sigma_{i+}\sigma_{i-}} \right) = (-1)^n\frac{d^n\sigma}{\vol \SL(2)}\; \sum_{\rho\in S_{n}}\frac{1}{\sigma_{+\rho(1)}\sigma_{\rho(1)\rho(2)}\cdots\sigma_{\rho(n)-}\sigma_{-+}} \,,
\]
where we employ the notation\, $\sigma_{ab}=\sigma_a-\sigma_b$\,, with $a,b\in\{1,\cdots,n,+,-\}$, and apply the $\SL(2)$ fixing
\[
\frac{d^n\sigma}{\vol \SL(2)} := (\sigma_{+-}\sigma_{-1}\sigma_{1+}) \prod_{i=2}^n d\sigma_i\,.
\]
In the strict degeneration to the nodal sphere ($q=0$), $dz$ acquires poles at $\sigma_\pm$, and this pair of punctures represents a node associated to the loop momentum. Now, the objects $g^{(w)}_{ij}$ used for building $\cI_n$ take the following form as $q\to0$:
\begin{equation}
\label{eq:glim}
 g^{(1)}_{ij} \to
 \pi i \,\frac{\sigma_{i+}\sigma_{j-}+\sigma_{i-}\sigma_{j+}}{\sigma_{ij}\sigma_{+-}}\,,
 \qquad  g^{(2K)}_{ij} \to -2 \,\zeta(2K) \,, \qquad g^{(2K+1)}_{ij} \to 0\,,
\end{equation}
for $K\geq1$. As a result of \eqref{eq:G2k}, we also have $\,G_{2K}(\tau) \to 2 \,\zeta(2K)\,$.

Putting all these ingredients together, and assuming that `closed cycles' of the objects $g^{(w)}_{ij}$ are excluded from the chiral integrand $\cI_n$ as mentioned in the previous subsection, we conclude that
\[
\label{eq:In0}
(-1)^n(2\pi i)^3\, \,\cI_n(\ell) \prod_{i=2}^n dz_i  \;\to\; \frac{d^n\sigma}{\vol \SL(2)}\; \sum_{\rho\in S_{n}}\frac{N(\rho(1),\rho(2),\cdots,\rho(n);\ell)}{\sigma_{+\rho(1)}\sigma_{\rho(1)\rho(2)}\cdots\sigma_{\rho(n)-}\sigma_{-+}}
\]
as $q\to0$.
The fact that the (nodal-)sphere differential form admits this expression is well understood from both the mathematics \cite{10.2969/jmsj/03920191,Brown:2019wna} and the physics \cite{Mafra:2011kj,Mafra:2011nv,Cachazo:2013iea,Geyer:2015bja,He:2015yua,Geyer:2015jch} literature, and is crucial for obtaining a standard loop integrand in field theory. The coefficients $N$ on top of the `Parke-Taylor' denominators are independent of the marked points $\sigma_a$, and turn out to be a set of master BCJ numerators, as we shall discuss.\footnote{The analogy between \eqref{eq:In0} and the associated object at tree level appearing in \eqref{eq:Atree} is obvious.}

\begin{figure}[t]
    \begin{align}
        \includegraphics[align=c,scale=0.45]{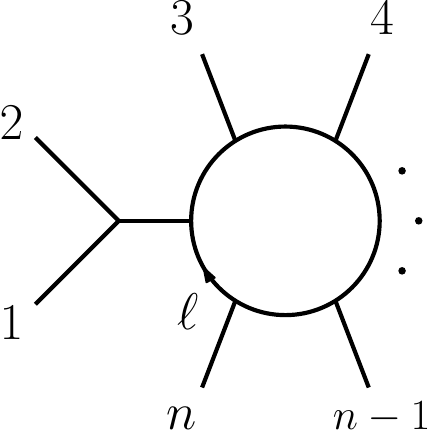} \quad = \quad
        \includegraphics[align=c,scale=0.45]{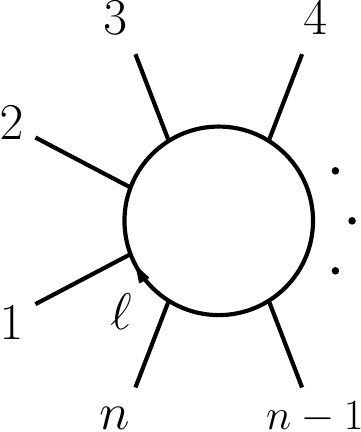} \quad - \quad
        \includegraphics[align=c,scale=0.45]{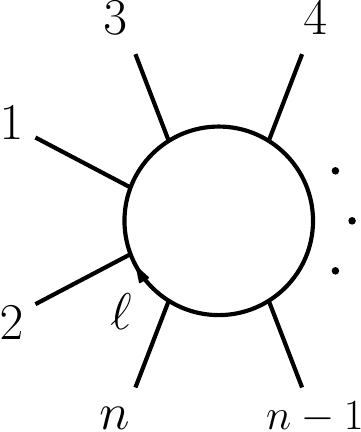} \nonumber
    \end{align}
\caption{Example of one-loop BCJ relation expressing a non-master numerator in terms of master numerators.}
\label{fig:N12}
\end{figure}

The BCJ construction of the field-theory loop integrand in terms of trivalent diagrams works as follows. The master numerators $N(\cdots;\ell)$ correspond to the $n$-gon diagrams, i.e.~those where the $n$ external lines meet the loop directly. See the diagrams on the right-hand side of figure~\ref{fig:N12}.
This figure illustrates how to construct the BCJ numerators for the other (non-master) trivalent diagrams via successive `commutators'; in particular, it illustrates the first of the following examples:
\begin{align}
\label{eq:Njacob}
N([1,2],3,\cdots,n;\ell)&=N(1,2,3,\cdots,n;\ell)-N(2,1,3,\cdots,n;\ell)\,, \nonumber \\
N([[1,2],3],\cdots,n;\ell)&=N([1,2],3,\cdots,n;\ell)-N(3,[1,2],\cdots,n;\ell)\,, \nonumber \\
N([1,2],[3,4],\cdots,n;\ell)&=N([1,2],3,4,\cdots,n;\ell)-N([1,2],4,3,\cdots,n;\ell)\,,
\end{align}
and so on. Given the master numerators, we can obtain the numerators for all trivalent diagrams via these Jacobi-type relations. The loop integrands in super-Yang-Mills and supergravity are given by the one-loop instance of the BCJ double copy \cite{Bern:2010ue}:
\begin{equation}
\label{eq:Aft}
 \mathcal{A}_n^{(1),\text{SYM}} = \int_{\mathbb R^D} d^D\ell\;\sum_{\gamma\in\Gamma_{n,3}^{(1)}} \frac{N_\gamma(\ell)\; c_\gamma}{D_\gamma} \,,
 \qquad 
 \mathcal{A}_n^{(1),\text{SUGRA}} = \int_{\mathbb R^D} d^D\ell\;\sum_{\gamma\in\Gamma_{n,3}^{(1)}} \frac{N_\gamma(\ell)\; \tilde N_\gamma(\ell)}{D_\gamma} \,.
\end{equation}
The sums are over the set $\Gamma_{n,3}^{(1)}$ of distinct trivalent one-loop diagrams. The factors $1/{D_\gamma}$ are the product of scalar propagators for each diagram, and the factors $c_\gamma$ in gauge theory are the colour factors. The latter trivially satisfy relations analogous to \eqref{eq:Njacob} by virtue of the Jacobi identity for the colour Lie algebra, hence the term `colour-kinematics duality'. We remind the reader that the expressions \eqref{eq:Aft} for super-Yang-Mills and supergravity are ---  in our 10D setting --- only expressions for the loop integrands, as the field-theory amplitudes themselves are not defined due to ultraviolet divergence. The fact that BCJ numerators indicate the existence of a `kinematic algebra' that mirrors the colour Lie algebra has been the subject of extensive work, e.g.~\cite{Mafra:2011kj,Monteiro:2011pc,Boels:2011mn,Broedel:2012rc,Boels:2012ew,Boels:2013bi,Monteiro:2013rya,Cheung:2016prv,Du:2017kpo,Fu:2018hpu,Chen:2019ywi,Reiterer:2019dys,Borsten:2021hua,Chen:2021chy,Cheung:2021zvb,Brandhuber:2021bsf,Ben-Shahar:2021doh,Ben-Shahar:2021zww,Bonezzi:2022yuh,Brandhuber:2022enp,Borsten:2022vtg,Ben-Shahar:2022ixa,Bonezzi:2023pox,Borsten:2023ned,Armstrong-Williams:2024icu,Chen:2024gkj,Bonezzi:2024fhd,Ben-Shahar:2024dju,Fu:2025jpp,Ben-Shahar:2025dci}, much of it connected to aspects of string theory. 

We mention now some properties of the BCJ numerators that will be important for us; see \cite{Bern:2019prr} for a comprehensive description of the BCJ story. As in figure~\ref{fig:N12}, the number of commutators reduces the number of legs along the loop, and, because we are dealing with maximal supersymmetry, the numerators vanish once we reach a triangular loop. A related feature of maximal supersymmetry is that the master numerators are polynomials of order $n-4$ in the loop momentum, and each commutator reduces the order of this polynomial by one order; e.g.~at five points, the master (pentagon) numerators are linear in $\ell$, while the box numerators (with one massive corner) obtained as in figure~\ref{fig:N12} are independent of $\ell$. These properties strongly constrain the numerators, in a manner that can also be seen as resulting from the superstring ansatz. For instance, the $\ell^{n-4}$ piece of the master numerators matches via \eqref{eq:In0} the $\ell^{n-4}$ piece of the chiral integrand $\cI_n(\ell)$, i.e.~the piece of $\cI_n(\ell)$ whose weight is fully carried by a power of $\ell$. It is important also to mention the automorphic properties of the master numerators, namely the reflection property,
\[
\label{eq:Nrefl}
N(1,2,3,\cdots,n-1,n;\ell) = (-1)^n N(n,n-1,\cdots,3,2,1;-\ell)\,,
\]
and the quasi-cyclic property,
\[
\label{eq:cyclic}
N(1,2,3,\cdots,n-1,n;\ell) = N(2,3,\cdots,n-1,n,1;\ell+p_1)\,.
\]
The latter property is hopefully clear from the middle diagram in figure~\ref{fig:N12}: the left- and right-hand sides differ only by which internal leg of the $n$-gon we choose to define the loop momentum. The colour counterparts of these properties (under the `colour-kinematics duality') are trivial: the colour factors of $n$-gons, $c^{a_1a_2\cdots a_n}:=f^{b_1a_1b_2}f^{b_2a_2b_3}\cdots f^{b_na_nb_1}$, are such that $c^{a_1a_2\cdots a_n}=(-1)^n c^{a_n\cdots a_2a_1}=c^{a_2\cdots a_na_1}$.

The type of $\alpha'\to 0$ manipulation leading from the string amplitudes to \eqref{eq:Aft} via the degeneration \eqref{eq:In0} is well-understood in the literature; see the recent work \cite{Balli:2024wje} for explicit examples. Following our logic of `inverting' the limit $\alpha'\to 0$, we will map directly pieces of the BCJ numerators to the coefficients of the superstring chiral integrand $\cI_n$. The automorphic properties of the numerators will allow us to decompose them in a manner that simplifies the formulation of this map. It is worth emphasising that the quasi-cyclic property \eqref{eq:cyclic} can be understood as following from the invariance under monodromy \eqref{eq:monodromy} of $\cI_n$. To see this, consider the monodromy for particle 1:
\[
\label{eq:monod1}
\ell_\mu \mapsto \ell_\mu - p_{1,\mu}\,, \qquad 
z_1\mapsto z_1+\tau\,, \qquad z_{i>1}\mapsto z_i\,.
\]
Effectively, this corresponds in the degeneration limit to, for instance,
\[
\frac{N(1,2,\cdots,n;\ell)}{\sigma_{+1}\sigma_{12}\sigma_{23}\cdots\sigma_{n-}\sigma_{-+}}
\quad \mapsto \quad
\frac{N(1,2,\cdots,n;\ell-p_1)}{\sigma_{+2}\sigma_{23}\cdots\sigma_{n1}\sigma_{1-}\sigma_{-+}}\,.
\]
Now, the quasi-cyclic property \eqref{eq:cyclic} simply identifies the latter with
\[
\frac{N(2,3,\cdots,n,1;\ell)}{\sigma_{+2}\sigma_{23}\cdots\sigma_{n1}\sigma_{1-}\sigma_{-+}}\,.
\]
All the `BCJ monodromies' described in \cite{Geyer:2024oeu} follow from the general relation \eqref{eq:cyclic}. It implies that $N(A,B;\ell)=N(B,A;\ell+p_A)$, where $A$ and $B$ are complementary sets of the external particles. This quasi-cyclic property of the numerators is essential for the recombination of the field-theory limit into the standard loop-integrand expressions \eqref{eq:Aft}.

We conclude this section with a comment for the readers who are familiar with the ambitwistor string story --- other readers may safely skip it. The chiral integrands that we construct in this paper can be imported into the ambitwistor string, and lead to loop integrands with standard `quadratic' propagators, as opposed to the `linear' propagators that appear generically from the one-loop scattering equations on the nodal sphere, e.g.~\cite{Geyer:2015jch,Geyer:2017ela}. This occurs because of the quasi-cyclic property of the numerators extracted via \eqref{eq:In0}. Consider the example of a bubble diagram (which easily generalises to higher-gons):
\begin{align}
\frac1{\ell^2} \left[ \frac{N(A,B;\ell)}{2\,\ell\cdot p_A+p_A^2} + \frac{N(B,A;\ell)}{-2\,\ell\cdot p_A+p_A^2} \right] & \; \stackrel{\text{shift}}{\longmapsto}\,  \frac{N(A,B;\ell)}{{\ell^2}(2\,\ell\cdot p_A+p_A^2)} + \frac{N(B,A;\ell+p_A)}{{(\ell+p_A)^2}(-2\,\ell\cdot p_A-p_A^2)}  \nonumber \\  & =\, \frac{N(A,B;\ell)}{{\ell^2}(2\,\ell\cdot p_A+p_A^2)} + \frac{N(A,B;\ell)}{{(\ell+p_A)^2}(-2\,\ell\cdot p_A-p_A^2)}  = \frac{N(A,B;\ell)}{ \ell^2(\ell+p_A)^2}\,.
\nonumber
\end{align}
In the first step, a loop-momentum shift $\ell\mapsto\ell+p_A$ was performed in the second term only; in the second step, the quasi-cyclic property was used; in the final step, a standard diagrammatic contribution was obtained. We note also that the quasi-cyclic property of the numerators follows from the monodromy invariance of a chiral integrand $\cI_n$ that presents no `closed cycles', which is not the case for `CHY Pfaffians'.


\section{A first look at the worldsheet basis}
\label{sec:elliptic}

In this section, we introduce the simplest building blocks of the superstring ansatz. We follow a different approach to our previous one-loop work \cite{Geyer:2024oeu}. There, at each multiplicity, we started with a general ansatz for the chiral integrand $\cI_n$ using the objects in the table~\ref{eq:tablew}, with weight $n-4$, and then imposed invariance under monodromy \eqref{eq:monodromy}, which in turn puts constraints on the coefficients of the ansatz. Here, instead, we will build on an idea of \cite{Mafra:2017ioj} (see also \cite{Rodriguez:2023qir,Zhang:2024yfp}) by working with an ansatz where each term is already monodromy-invariant on its own. In the language of \cite{Mafra:2017ioj}, each term is, therefore, a `generalised elliptic integrand': while elliptic refers to doubly-periodic, generalised elliptic refers to doubly-periodic once the loop momentum is also shifted appropriately \eqref{eq:monodromy}.

Consider weight 1, which will be sufficient for the 5-point amplitude. The object
\begin{equation}
\label{eq:L}
    L_{\mu} := 2\pi i \, \ell_{\mu} + \sum_{i=2}^{n} p_{i,\mu} \, \egg{1,i}
\end{equation}
is easily checked, using momentum conservation, to be invariant under monodromy \eqref{eq:monodromy}. This is achieved by sacrificing permutation symmetry in the external particles: we choose particle 1 to take a privileged role in the worldsheet basis. It turns out to be extremely convenient. One can check that, at weight 1, the space of `generalised elliptic integrands' is spanned by $L_\mu$ and by the objects
\begin{equation}
\label{eq:E1ij}
    V_{1|i,j} := \egg{1,i} + \egg{i,j} - \egg{1,j} \,, \qquad 2 \leqslant i < j \leqslant n \,,
\end{equation} 
which are also monodromy-invariant.\footnote{For instance, \,$\displaystyle 2\pi i \, \ell_{\mu} + \sum_{i=1}^{n-1} p_{i,\mu} \, \egg{n,i} = L_\mu - \sum_{i=2}^{n-1} p_{i,\mu}\, V_{1|j,n}$\,.}

At higher weight, the story is more involved. We may consider products of $L_\mu$ and/or $V_{1|i,j}$, e.g.~$V_{1|i,j} V_{1|k,l}$ at weight 2; and we may also consider other objects, such as 
$$
V_{1|i,j,k} := \, g^{(2)}_{1,i} + g^{(2)}_{i,j} + g^{(2)}_{j,k} + g^{(2)}_{k,1} + g^{(1)}_{1,i} g^{(1)}_{i,j} + g^{(1)}_{1,i} g^{(1)}_{j,k} + g^{(1)}_{1,i} g^{(1)}_{k,1} + g^{(1)}_{i,j} g^{(1)}_{j,k} + g^{(1)}_{i,j} g^{(1)}_{k,1} + g^{(1)}_{j,k} g^{(1)}_{k,1}\,,
$$
again at weight 2.

The elements of the worldsheet basis that do not involve the loop momentum are easy to describe at $n$ points, and in fact this was already achieved in \cite{Mafra:2018pll}. The objects $V$ above generalise as
\[
\label{eq:VFF}
V_{1|i_1,i_2,\cdots,i_s}:= F(z_1-z_{i_1},\eta,\tau) F(z_{i_1}-z_{i_2},\eta,\tau) F(z_{i_2}-z_{i_3},\eta,\tau) \cdots
F(z_{i_s}-z_1,\eta,\tau)\Big|_{\eta^{-2}}  \,,
\]
where the $i_\text{a}>1$ label distinct punctures.\footnote{This makes it manifest that the $V$'s are elliptic, because the quasi-doubly-periodicity of $F$ turns into doubly-periodicity of this cyclic product. These objects made an earlier appearance in \cite{Dolan:2007eh}.}
The Kronecker-Eisenstein series \eqref{eq:F} can be rewritten with a substitution rule:
\[
F(z,\eta,\tau)=\frac1{\eta}\, e^{\eta\, g^{(1)}(z,\tau)}\Big|_{ \big(g^{(1)}(z,\tau)\big)^{m} \,\mapsto\, m! \, g^{(m)}(z,\tau)}\,.
\]
It follows that
\begin{equation}
    \label{eq:Vs}
    \boxed{ 
    \begin{aligned}
       V_{1|i_1,i_2,\cdots,i_s} = \, \frac{1}{(s-1)!} \left.\left( \egg{1,i_1} + \egg{i_1,i_2} + \cdots + \egg{i_{s-1},i_{s}} + \egg{i_s,1} \right)^{s-1} \right|_{ (\egg{a,b})^{m} \,\mapsto\, m! \, \eg{m}{a,b}} \,.
    \end{aligned}
    }
\end{equation}
These objects have weight $s-1$.
In fact, we also include the weight-0 cases $V_{1|i}:=1$. The substitution rule prevents the appearance of closed cycles, namely instances of $(\egg{a,b})^m$, which have a pole of order $m$ as $z_a\to z_b$.\footnote{We recall that $\egg{a,b}$ has a single pole of unit residue as $z_a\to z_b$.} The elements of the worldsheet basis that do not involve the loop momentum are simply
\[
\label{eq:prodE}
V_{1|i_1,i_2,\cdots,i_{s_1}} V_{1|j_1,j_2,\cdots,j_{s_2}}
V_{1|k_1,k_2,\cdots,k_{s_3}}\,,
\]
where the indices $i_\text{a},j_\text{a},k_\text{a}$ are all distinct and, together with 1, make up the $n$ punctures, so that $1+\sum_{r=1}^3 s_r = n$. This identity implies that the weight of \eqref{eq:prodE} is $\sum_{r=1}^3 (s_r-1) = n-4$, as required for a worldsheet basis element of the chiral integrand $\cI_n$. There is an additional constraint on these basis elements:
\[
\label{eq:indV}
i_1<i_\text{a}\,,\quad j_1<j_\text{a}\,, \quad k_1<k_\text{a}\,, \qquad \text{for}\; \text{a}>1\,.
\]
Objects of the form \eqref{eq:prodE} that do not obey this condition can be obtained from those that do, using the Fay identities that follow from \eqref{eq:Fay}.\footnote{For instance, taking $1<i<j<k$, the Fay identities imply that \,$V_{1|j,i,k} = - V_{1|i,k,j} - V_{1|i,j,k}$\,. Such linear relations generalise as a `shuffle' symmetry of the $V$'s \cite{Mafra:2018pll}, which allows to choose a basis of obeying \eqref{eq:indV}.} 

Let us consider some examples:
\begin{itemize}
    \item $n=4$:\, 1 basis element given by\, $V_{1|2}V_{1|3}V_{1|4}=1$\,, so there is no dependence on the punctures;
    \item $n=5$:\, 6 basis elements given by\, $V_{1|i,j}\prod_{k\neq 1,i,j}V_{1|k}=V_{1|i,j}$\, with $1<i<j$;
    \item $n=6$:\, 15 basis elements given by  \,$V_{1|i,j}V_{1|k,l}$\, with distinct indices obeying $1<i<j,k$ and $k<l$; and 20 basis elements given by \,$V_{1|i,j,k}$\, with distinct indices obeying $1<i<j,k$.
\end{itemize}
We will see later that the kinematic coefficients corresponding to these basis elements are BCJ numerators of box diagrams where particle 1 is at one corner, and each of the other three corners is associated to one of the three $V$'s in \eqref{eq:prodE}.

The elements of the worldsheet basis that involve the loop momentum are not as straightforward. Naively, one may extend the basis elements \eqref{eq:prodE} to include the loop momentum as
\[
L_{\mu_1} L_{\mu_2} \cdots L_{\mu_s} \times (\text{product of $V$'s of weight $n-4-s$})\,.
\]
Already at weight 2, however, the objects  $L_\mu L_\nu$ and $L_\mu V_{1|i,j}$ include terms of the type $(\egg{1,i})^2$. At higher weights, more general closed cycles would appear. One of the main results in this paper is an $n$-point regularisation rule that deals with basis elements associated to any power of $L_\mu$, which also incorporates at higher points the holomorphic Eisenstein series. We leave the description of the complete worldsheet basis, including the definition of the regularisation rule, to section~\ref{sec:npt}. By then, the reader will have encountered illustrative examples for low multiplicity.

At 4 and 5 points, the proof of completeness of our worldsheet basis is straightforward. We leave the 6-point basis for section~\ref{sec:6pt}, but there is one comment we may make here. In  \cite{Mafra:2018pll}, an expression for the 6-point chiral integrand was derived using the pure-spinor formalism, including only one worldsheet function that lives outside the function space we consider. In the open string integral, this function reads:
\begin{equation}
    E_{1|2|3,4,5,6} := \frac{1}{2\alpha'} \partial_1 \egg{12} + p_1 \cdot p_2\, (\egg{12})^2 - 2 \, p_1 \cdot p_2 \,\eg{2}{12} \,,
\end{equation}
while for the closed-string integrand it is obtained by the replacement $\alpha' \mapsto \alpha'/4$. Note both the $\alpha'$ dependence and the closed cycle. However, it was shown in~\cite{Balli:2024wje} that this function can be re-expressed, up to integration by parts, in terms of objects that correspond to our basis:
\begin{equation}\label{eq:6ptibp}
    E_{1|2|3,4,5,6} \simeq \frac{1}{p_1\cdot p_2} \left( p_{1}^{\mu} p_2^{\nu} \, \reg\left\lbrack L_{\mu} L_{\nu} \right\rbrack + \sum_{i=3}^{n} p_{1}^{\mu} (p_2 \cdot p_i) \, \reg\left\lbrack L_{\mu} V_{1|2,i} \right\rbrack \right) \,,
\end{equation}
where \reg{} denotes the regularisation whose definition at weight 2 will be provided in section~\ref{sec:6pt}. We discuss this further in appendix~\ref{app:ibp7pt}, where we also list the analogous integration-by-parts relations that exclude closed cycles at 7 points. This supports our conjecture that the all the generalised elliptic integrands with closed cycles can be excluded at any multiplicity. Notice also that the $\alpha'$ dependence has dropped out in \eqref{eq:6ptibp}.
We will leave the study of these integration-by-parts relations at higher multiplicity to future work. As we mentioned, we will present the $n$-point prescription for \reg{} in section~\ref{sec:npt}. Moreover, in appendix~\ref{app:proof}, we will sketch a proof of our worldsheet basis under the assumption of our conjecture of no closed cycles. 

In the following sections, we will present examples of our worldsheet basis starting at low multiplicity. As we write the chiral integrand in this basis, the kinematical coefficients will turn out to have a remarkably simple relation to the field-theory BCJ numerators.


\section{4-point chiral integrand}
\label{sec:4pt}

We start with the 4-point chiral integrand $\cI_4$, which is trivial from the worldsheet perspective. As we have already noted, its weight $n-4=0$ means that it is constant on the torus, so it depends only on the external particle data. Via \eqref{eq:In0}, we can identify it directly with the BCJ numerator
\[
\cI_4 = N(1234;\ell) = N_4 \,.
\]
The last equality denotes that the numerator is independent of the loop momentum (as otherwise the expression would carry weight) and is also independent of the particle ordering (as otherwise there would be triangle numerators, which are absent for maximal supersymmetry). For NS external states, the expression for the chiral integrand is a long-known result \cite{Green:1981yb,Schwarz:1982jn,Green:1982sw}.\footnote{Explicitly, \,$N_4=\tr(f_1 f_2 f_3 f_4)
- \frac{1}{4} \tr(f_1 f_2 ) \tr(f_3 f_4) + \text{cyc}(2,3,4)$\,, where\, $f_i^{\mu\nu} = p_i^\mu \epsilon_i^\nu - \epsilon_i^\mu p_i^\nu$\,, and `cyc' denotes a sum over cyclic permutations.}


\section{5-point chiral integrand}
\label{sec:5pt}

The chiral integrand (weight: $n-4=1$) can be written as
\begin{align}
\label{eq:I5comp}
\cI_5 \, &= 2\pi i\;C_5^\mu \,\ell_\mu + \sum_{i<j}\, C_{5,ij}\, g^{(1)}_{ij} \nonumber \\
& = C_5^\mu\, L_{\mu} + \sum_{2 \leqslant i < j \leqslant 5} C_{5,ij} \, V_{1|i,j}\,.
\end{align}
In the first line, we wrote it as in \cite{Geyer:2024oeu}, whereas in the second line we wrote it in the manifestly monodromy-invariant basis from section~\ref{sec:elliptic}. Notice that the coefficients $C_{5,1j}$ are absent in the second line, which reflects the fact that the `basis' in the first line is overcomplete. To match the two expressions, we need to exploit the relation among the coefficients that follows from the monodromy invariance \eqref{eq:monodromy} of the first line:
\[
\label{eq:5ptmonod}
C_5^\mu\, {p_{i\mu}} + \sum_{j\neq i} \,C_{5,ij} =0 \,.
\]
Whichever expression we choose in \eqref{eq:I5comp}, we can use the degeneration formula \eqref{eq:In0} to read off the map between the kinematic coefficients of the chiral integrand and the field-theory BCJ numerators:
\[
C_5^\mu = N(\cdots;\ell)\big|_{\ell_\mu}=: N_5^\mu\,, \qquad C_{5,ij} = - N(\cdots[i,j]\cdots;\ell)=:-N_5([i,j])\,,
\label{eq:C5N}
\]
with the BCJ numerators taking the form
\begin{align}
\label{eq:5ptNgen}
N(12345;\ell) \,& = N(\cdots;\ell)\big|_{\ell_\mu} \, \ell_{\mu} + N(12345;\ell)\big|_{\ell^0} \nonumber \\ 
&=
N_5^\mu\, \ell_{\mu} +\frac1{2}\, \sum_{i<j}\,N_5([i,j])\,.
\end{align}
As mentioned in section~\ref{subsec:FTlimit}, the structure of the BCJ numerators is such that the master numerators are polynomials of order $n-4=1$ in $\ell$, and such that each commutator reduces the order in $\ell$ by one. Hence, the use of the ellipsis above denotes that the ordering of the particles that are not explicitly included is irrelevant. The notation $(\cdot)|_{\ell_\mu}$ means that we extract the coefficient of $\ell_\mu$ in the numerator; similarly, $(\cdot)|_{\ell^0}$ means that we extract the numerator evaluated at $\ell=0$. The decomposition \eqref{eq:5ptNgen} of the numerators could have been anticipated from their general properties.\footnote{Let us see how we could have anticipated \eqref{eq:5ptNgen}. Box numerators, obtained as a `commutator' of pentagon numerators, have only a $\ell^0$ piece, so the part $\ell^1$ of the pentagon numerators is independent of ordering. In addition, as triangle numerators must vanish, any `double or higher commutator' of pentagon numerators is absent in the decomposition. Finally, the reflection symmetry \eqref{eq:Nrefl} excludes any ordering-independent $\ell^0$ piece of the pentagon numerators.}
We note also that the condition of monodromy invariance of the chiral integrand \eqref{eq:5ptmonod} matches the quasi-cyclic property \eqref{eq:cyclic} of the numerators:
\[
\label{eq:5ptquasicyc}
0 = N(i,\rho_4;\ell) -N(\rho_4, i;\ell+p_i) = -N_5^\mu\, {p_{i\mu}} + \sum_{j\neq i} \,N_5([i,j]) \,,
\]
where $\rho_4$ is any ordering of the four not-$i$ labels. An explicit expression for the 5-point 10D BCJ numerators can be found in \cite{Edison:2022jln}.\footnote{The BCJ numerators are not gauge-invariant, but a pure gauge $\cI_5$, together with the Koba-Nielsen factor, leads to a total derivative in moduli space. See e.g.~(9.11) of \cite{Geyer:2024oeu}.}

With the map \eqref{eq:C5N}, we can finally write the chiral integrand using a manifestly monodromy-invariant basis, whose coefficients are pieces of BCJ numerators:
\begin{equation}
\label{eq:I5ansatz}
\boxed{\,
    \cI_{5} = N_5^\mu\, L_{\mu} - \sum_{2 \leqslant i < j \leqslant 5} N_5([i,j]) \, V_{1|i,j} \,.
    }
\end{equation}
In fact, this 5-point expression was already noted in \cite{Mafra:2017ioj}.
We emphasise here that the kinematic coefficients can be determined by any valid set of BCJ numerators, so their derivation is not reliant on the use of any worldsheet formalism (e.g.~RNS or pure spinor). We start to see in this expression a feature that was also first noticed in \cite{Mafra:2017ioj}: that there is a type of duality between the worldsheet functions and the kinematic coefficients. Our basis of monodromy-invariant worldsheet functions is designed to make this duality manifest at higher multiplicity.


\section{6-point chiral integrand}
\label{sec:6pt}

The story becomes more interesting starting at 6 points. At weight $2=n-4$, we expect products of weight-1 objects in our basis, which we will need to regularise to avoid closed cycles. We will also describe a new constraint on the field theory that arises from modularity on the worldsheet.

A simple expression for the 6-point chiral integrand in terms of a monodromy-invariant worldsheet basis is
\begin{equation}
    \boxed{ \,
    \begin{aligned}
        \cI_{6} = &\, N_6^{\mu\nu} \, \reg\left\lbrack L_{\mu} L_{\nu} \right\rbrack - 
        \sum_{2 \leqslant i<j \leqslant 6} N_6^\mu([i,j]) \, \reg \left\lbrack L_{\mu} \, V_{1|i,j} \right\rbrack \\
        & + \sum_{ \substack{2 \leq i < j , k \leq 6 \\  j \neq k } } N_6([[i,j],k])\, V_{1|i,j,k} + \sum_{ \substack{ 2 \leqslant i<j \leqslant 6 \\ 2 \leqslant i<k<l \leqslant 6 \\ j \neq k,l } } N_6([i,j],[k,l]) \, V_{1|i,j} V_{1|k,l} \,. 
    \end{aligned} \,
    }
    \label{eq:6ptreg}
\end{equation}
The kinematic coefficients obtained from BCJ numerators are $\ell$-independent, with
\[
N_6^{\mu\nu} := N(\cdots;\ell)\big|_{\ell_\mu\ell_\nu}\,,
\qquad
N_6^\mu([i,j]) := N(\cdots[i,j]\cdots;\ell)\big|_{\ell_\mu}\,,
\]
where the ellipses indicate that the ordering of the particle labels not explicitly presented is irrelevant.\footnote{Our notation is:\; $N(\rho;\ell)\big|_{\ell_{\mu_1}\ell_{\mu_2}\cdots\ell_{\mu_p}} = \frac1{p!}\frac{\partial^p}{\partial\ell_{\mu_1}\partial\ell_{\mu_2}\cdots\partial\ell_{\mu_p}}\, N(\rho;\ell)\Big|_{\ell=0}\,. $}
The expression \eqref{eq:6ptreg} matches, but greatly simplifies, an analogous expression in \cite{Geyer:2024oeu}. Let us discuss the worldsheet basis used here, which has weight $n-4=2$. We have already encountered in section~\ref{sec:elliptic} the basis elements appearing in the second line. The worldsheet functions in the first line involve products of weight-1 objects $L_{\mu}$ and $V_{1|i,j}$ which give rise to closed cycles. We will denote by `\reg' a linear operation that regularises the closed cycles, which at weight 2 is realised via the replacement rule
\begin{equation}\label{eq:defreg6pt}
    \begin{aligned}
        \reg\left\lbrack g_{i,j}^{(1)} g_{k,l}^{(1)} \right\rbrack &:= \egg{i,j}\egg{k,l} \,, \qquad \{ i,j \} \neq \{ k,l \} \,, \\
        \reg\left\lbrack \left( g_{i,j}^{(1)} \right)^2 \right\rbrack &:= \left( \egg{i,j} \right)^2 + \partial_i g^{(1)}_{i,j} + G_2(\tau) = 2\eg{2}{i,j} \,.
    \end{aligned}
\end{equation}
The last equality follows from \eqref{eq:g1g2}. A significant part of our work at higher multiplicity will be to generalise this replacement rule acting on products of $L$'s and $V$'s. In a later section, we will motivate the rule and provide a general prescription.

Now, let us focus on the kinematic coefficients in \eqref{eq:6ptreg}. We can see a general $n$-point structure arising. The coefficient associated to the (regularised) product of $n-4$ $L$'s is the coefficient of $\ell^{n-4}$ in the master BCJ numerators, which is independent of the ordering. The coefficients associated to the (regularised) products of $n-5$ $L$'s and a $V_{1|ij}$ are the coefficients of $\ell^{n-5}$ in the sub-master numerators (one commutator, independent of other ordering). The coefficients associated to the (regularised) products of $n-6$ $L$'s (here, $L^0$) together with weight-2 products of $V$'s are the coefficients of $\ell^{n-6}$ in the sub-sub-master numerators (two commutators, independent of other ordering). This is as far as it gets at 6 points, but the pattern is clear. We will discuss later novelties arising at 8 and 16 points, associated to modular forms.

As observed in \cite{Geyer:2024oeu}, there is a puzzle, however. Let us try to write a decomposition of the master numerators analogous to what we presented at 5 points in \eqref{eq:5ptNgen}, following solely from generic properties of one-loop BCJ numerators in maximally supersymmetric theories:
\begin{align}
N(123456;\ell) \, & =\, N(123456;\ell)\big|_{\ell_\mu\ell_\nu}\; \ell_{\mu}\ell_\nu +
N(123456;\ell)\big|_{\ell_\mu}\; \ell_\mu + N(123456;\ell)\big|_{\ell^0} \nonumber \\
& =\, N_6^{\mu\nu}\, \ell_{\mu}\ell_\nu +\frac1{2}\, \sum_{i<j}\, N_6^\mu([i,j])\, \ell_\mu  + N(123456;\ell)\big|_{\ell^0}
\,.
\end{align}
The first two parts in each line follow a similar reasoning as at 5 points.\footnote{The piece quadratic in $\ell$ is independent of the ordering, because the pentagon numerators have are at most linear in $\ell$ at 6 points. The piece linear in $\ell$ cannot have a permutation symmetric piece, due to the reflection symmetry \eqref{eq:Nrefl}.} The last part, which is $\ell$-independent, is decomposed as\footnote{Triple commutators must vanish, because the numerators of triangle diagrams vanish. Moreover, a single commutator in the $\ell^0$ piece is excluded by the reflection symmetry.}
\begin{align}
N(123456;\ell)\big|_{\ell^0} = &\; \frac1{6}\sum_{i<j<k} \left(N([[i,j],k]+N([i,[j,k]]\right)
+ \frac{1}{4} \sum_{ \substack{ i<j\,;\,k<l \\ i<k\,;\,j\neq k,l } } N([i,j],[k,l]) \nonumber \\
&\, + \frac{1}{6!} \sum_{\rho \in S_6} N(\rho\,;\ell) \big|_{\ell^0}\,.
\end{align}
The last term indicates that, at 6 points, we have an $\ell^0$ piece that is independent of particle ordering.\footnote{At 5 points, and in fact at any odd $n$, this is excluded by the  reflection symmetry \eqref{eq:Nrefl}.} This piece drops out of the quasi-cyclic relation of the numerators \eqref{eq:cyclic} so the standard BCJ properties do not constrain it in any way. Importantly, this piece does not contribute to the superstring chiral integrand \eqref{eq:6ptreg}. Reversing the logic, the numerators for maximally supersymmetry super-Yang-Mills/supergravity can be derived from the chiral integrand via degeneration; hence, this piece is not an independent piece of the BCJ numerators for these particular theories --- it is determined by the other pieces. There are different ways of expressing this, because various pieces of the master numerators are related by the quasi-cyclic property. The simplest expression is
\begin{equation}
\label{eq:6ptidG2}
    \boxed{
        \frac{1}{6!} \sum_{\rho \in S_6} N(\rho\,;\ell) \big|_{\ell = 0} = \frac{1}{12} \, N_{6,\mu\nu}  \sum_{i=1}^6  p_i^\mu p_i^\nu \,.
    }
\end{equation}
Any valid set of BCJ numerators for the maximal supersymmetric theories, including in the 4D case obtained by dimensional reduction, must obey this quite non-trivial property.

There is another perspective to understand this constraint. At weight 2, there is also the function $G_2(\tau)$ defined in \eqref{eq:G2}, which has a non-vanishing limit as $\tau\to i \infty$. However, as we noted then, $G_2(\tau)$ is \textit{not} a modular form, and thus has to be excluded from the ansatz due to modularity. If we naively add to $\cI_6$ in \eqref{eq:6ptreg} the term $C_{6,G_2}\,G_2(\tau)$, where $C_{6,G_2}$ is a kinematic coefficient, this coefficient would contribute to the BCJ numerators extracted from the degeneration formula \eqref{eq:In0}. Following our logic of inverting this dependence to determine $C_{6,G_2}$ in terms of the numerators, we would find an expression for $C_{6,G_2}$ that vanishes precisely when the relation \eqref{eq:6ptidG2} holds. That is, requiring the would-be coefficient of $G_2(\tau)$ in $\cI_6$ to vanish leads to the above identity. In this sense, this field-theory relation results from the modularity of the superstring.

For a sanity check, we have verified the identity \eqref{eq:6ptidG2} using the explicit one-loop BCJ numerators determined in~\cite{Edison:2022jln}. We note that these were {\it not} determined starting from string theory, but rather from a purely field-theory construction. This raises a natural question: what is the physical relevance of this identity from a field-theory perspective? Is it a new symmetry of maximally supersymmetric field theory? We leave an investigation of this question to future work.

\section{7-point chiral integrand}
\label{sec:7pt}

The expression for the 7-point chiral integrand in terms of a monodromy-invariant worldsheet basis is
\begin{equation}
    \boxed{
    \begin{aligned}
        \cI_{7} =&\; N_7^{\mu\nu\rho} \, \reg \left\lbrack L_{\mu} L_{\nu} L_{\rho} \, \right\rbrack - \sum_{2 \leqslant i<j \leqslant 7} N^{\mu\nu}_{7}([i,j]) \, \reg \left\lbrack L_{\mu} L_{\nu} V_{1|i,j} \right\rbrack \\
        & + \sum_{2\leqslant i<j,k \leqslant 7 } N^{\mu}_{7}([[i,j],k]) \, \reg \left\lbrack L_{\mu} V_{1|i,j,k} \right\rbrack 
        + \sum_{ \substack{ 2 \leqslant i<j \leqslant 7 \\ 2 \leqslant k<l \leqslant 7 \\ i<k, j\neq l } } N^{\mu}_{7}([i,j],[k,l]) \, \reg\left\lbrack L_{\mu} V_{1|i,j} V_{1|k,l} \right\rbrack \\
        & - \sum_{ \substack{ 2 \leqslant i < j,k,l \\ j,k,l \, \text{dist} } } N_7([[[i,j],k],l]) \, V_{1|i,j,k,l} \,
        - \sum_{ \substack{ 2 \leqslant i < j,k \leqslant 7 \\ 2 \leqslant l<m \leqslant 7 \\ i,j,k,l,m \, \text{dist} } } N_7([[i,j],k], [l,m]) \, V_{1|i,j,k} V_{1|l,m} \\
        & - \sum_{ \substack{ 2 \leqslant i<j \leqslant 7 \\ 2 \leqslant i< k<l \leqslant 7 \\ 2\leqslant i<k< m<o \leqslant 7 \\ j,k,l,m,o \, \text{dist} } } N_7([i,j], [k,l], [m,o]) \, V_{1|i,j} V_{1|k,l} V_{1|m,o} \,,
    \end{aligned}
    }\label{eq:7ptreg}
\end{equation}
where `dist' means the indices listed are all distinct. We denote
\begin{align}
N_7^{\mu\nu\rho} := N(\cdots;\ell)\big|_{\ell_\mu\ell_\nu\ell_\rho}\,,&
\qquad
N_7^{\mu\nu}([i,j]) := N(\cdots[i,j]\cdots;\ell)\big|_{\ell_\mu\ell_\nu}\,, \\
N^{\mu}_{7}([[i,j],k]) := N(\cdots[[i,j],k]\cdots;\ell)\big|_{\ell_\mu}\,,&
\qquad
N^{\mu}_{7}([i,j],[k,l]) := N(\cdots[i,j]\cdots[k,l]\cdots;\ell)\big|_{\ell_\mu}\,.  \nonumber
\end{align}
As always, the loop momentum appears in the chiral integrand only via the worldsheet basis, not via the kinematic coefficients obtained from BCJ numerators. We define the regularisation of worldsheet functions at weight three as follows:
\begin{equation}\label{eq:7ptregrule}
    \begin{aligned}
        \reg \left\lbrack \left( g_{i,j}^{(1)} \right)^3 \right\rbrack &:= 6 \eg{3}{i,j} \,, \\
        \reg \left\lbrack \eg{2}{i,j} \eg{1}{i,j} \right\rbrack &:= 3 \eg{3}{i,j} \,, \\
        \reg \left\lbrack \left( \eg{1}{i,j} \right)^2 \eg{1}{k,l} \right\rbrack &:= 2 \eg{2}{i,j} \egg{k,l} \,, \\
        \reg \left\lbrack \egg{i,j} \egg{j,k} \egg{i,k} \right\rbrack &:= \egg{i,j} \eg{2}{i,k} + \egg{j,k} \eg{2}{i,k} - \eg{2}{i,j} \egg{i,k} - \eg{2}{j,k} \egg{i,k} - 3 \eg{3}{i,k} \,, \\
        \reg \left\lbrack \egg{i,j} \egg{k,l} \egg{m,n} \right\rbrack &:= \egg{i,j} \egg{k,l} \egg{m,n} \,.
    \end{aligned}
\end{equation}
Again, we leave the general prescription for a later section.\footnote{Some properties are not immediately manifest, e.g.~in the fourth line, the expression on the right-hand side is cyclically invariant due to Fay relations.} We note that \eqref{eq:7ptreg} is much  simpler than our previous 7-point result in~\cite{Geyer:2024oeu}, which is due to the choice of worldsheet basis.

Similarly to 6 points, there are identities like \eqref{eq:6ptidG2} among the BCJ numerators. In fact, there are now two types of identities:\footnote{
Note that the objects $N_n^{\mu_1\cdots \mu_s}$ are always fully symmetric in the spacetime indices.}
\begin{equation}
    \label{eq:7ptidG2}
    \boxed{
    \begin{aligned}
        & \frac{1}{7!} \sum_{\rho\in S_7} N(\rho; \ell) \big|_{\ell^{1}} = \frac{1}{4} \, N_7^{\mu\nu\rho} \;\ell_{\mu} \sum_{i=1}^{7} p_{i,\nu} p_{i,\rho} \,, \quad \forall \ell\,, \\
        & \frac{1}{6!} \sum_{\rho\in S_{6;[i,j]} } \!\! N(\rho; \ell) \big|_{\ell^{0}} = \frac{1}{12} \, N_7^{\mu\nu}([i,j]) 
        \left( 2 p_{i,\mu} p_{j,\nu} + \sum_{k=1}^{7} p_{k,\mu} p_{k,\nu} \right) \,, \\
    \end{aligned}
    }
\end{equation}
where $S_{6;[i,j]}$ in the second line denotes the 6!~permutations among the 5 particles $\notin \{i,j\}$ and the massive corner $[i,j]$. The first line above determines the $\ell^{1}$ piece of the fully symmetric part of the master BCJ numerators, while the second line determines the $\ell^{0}$ piece of the hexagon numerators after symmetrising over the 6 corners. On the one hand, neither of these pieces contributes to the coefficients of the chiral integrand \eqref{eq:7ptreg}. On the other hand, it is of course possible to derive the pieces from those coefficients via the degeneration formula \eqref{eq:In0}, which is how \eqref{eq:7ptidG2} is obtained, up to simplifications allowed by the quasi-cyclic property of the numerators.

At 6 points, we commented on the fact that the identity \eqref{eq:6ptidG2} could be understood as the vanishing, required by modularity, of the would-be kinematic coefficient of $G_2(\tau)$ in the chiral integrand $\cI_6$. Analogously, at 7 points the identities \eqref{eq:7ptidG2} can be understood as the vanishing of the kinematic coefficients $C^\mu_{7,G_2L}$ and $C_{7,G_2 V_{1|ij}}$, respectively, if we add to the chiral integrand $\cI_7$ the following term:
\[
G_2(\tau) \left( C^\mu_{7,G_2L}\,L_\mu + \sum_{2 \leqslant i<j \leqslant 7} C_{7,G_2 V_{1|ij}} \, V_{1|ij} \right)\,.
\]
Notice that there are only $\binom{6}{2}=15$ coefficients $C_{7,G_2 V_{1|ij}}$ in this expression, while there are $\binom{7}{2}=21$ identities in the second line of \eqref{eq:7ptidG2}. This means that, among those 21 identities, only 15 are independent: those where $i=1$  in \eqref{eq:7ptidG2} (taking $i<j$) follow from the others via relations among BCJ numerators, namely the Jacobi relations and the quasi-cyclic property \eqref{eq:cyclic}.

\section{8-point chiral integrand}
\label{sec:8pt}

At 8 points, the pattern seen so far is extended, but there is an important novelty. Since $\cI_8$ has weight $n-4=4$, an element of the holomorphic Eisenstein series is now allowed in the chiral integrand, namely $G_4(\tau)$. In the field-theory degeneration $\tau\to i\infty$, we have $G_4(\tau)\to 2\zeta(4)$. Because $G_4(\tau)$ is non-vanishing in this limit, we will be able to explicitly fix the corresponding coefficient in the chiral integrand in terms of field-theory BCJ numerators. This shows that the appearance of the holomorphic Eisenstein series is not an obstacle to fully fixing the superstring worldsheet correlator in terms of the field-theory limit --- at least at 8 points. We leave the higher-point discussion to the following section. The chiral integrand reads

\begin{equation}
    \boxed{
    \begin{aligned}
        & \cI_{8} = \, N_8^{\mu\nu\rho\sigma}\, \reg \left\lbrack L_{\mu} L_{\nu} L_{\rho} L_{\sigma} \, \right\rbrack - \sum_{2 \leqslant i<j \leqslant 7} N^{\mu\nu\rho}_{8}([i,j]) \, \reg \left\lbrack L_{\mu} L_{\nu} L_{\rho} V_{1|i,j} \right\rbrack \\
        & + \sum_{2\leqslant i<j,k \leqslant 8 } N^{\mu\nu}_{8}([[i,j],k]) \, \reg \left\lbrack L_{\mu} L_{\nu} V_{1|i,j,k} \right\rbrack 
        + \sum_{ \substack{ 2 \leqslant i<j \leqslant 8 \\ 2 \leqslant k<l \leqslant 8 \\ i<k, j\neq l } } N^{\mu\nu}_{8}([i,j],[k,l]) \, \reg\left\lbrack L_{\mu} L_{\nu} V_{1|i,j} V_{1|k,l} \right\rbrack \\
        & - \sum_{ \substack{ 2 \leqslant i < j,k,l \\ j,k,l \, \text{dist} } } N_8^{\mu}([[[i,j],k],l]) \, \reg \left\lbrack L_{\mu} V_{1|i,j,k,l} \right\rbrack \,
        - \sum_{ \substack{ 2 \leqslant i < j,k \leqslant 8 \\ 2 \leqslant l<m \leqslant 8 \\ i,j,k,l,m \, \text{dist} } } N_8^{\mu}([[i,j],k], [l,m]) \, \reg \left\lbrack L_{\mu} V_{1|i,j,k} V_{1|l,m} \right\rbrack \\
        & - \sum_{ \substack{ 2 \leqslant i<j \leqslant 8 \\ 2 \leqslant i< k<l \leqslant 8 \\ 2\leqslant i<k< m<n \leqslant 8 \\ j,k,l,m,o \, \text{dist} } } N_8^{\mu}([i,j], [k,l], [m,o]) \, \reg \left\lbrack L_{\mu} V_{1|i,j} V_{1|k,l} V_{1|m,o} \right\rbrack \\
        & + \sum_{ \substack{ 2 \leqslant i<j,k,l,m \leqslant 8 \\ j,k,l,m \, \text{dist} }} N_{8}([[[[i,j],k],l],m]) \, V_{1|i,j,k,l,m} \\
        & + \sum_{ \substack{2 \leqslant i<j,k,l \leqslant 8 \\ 2 \leqslant m<o \leqslant 8 \\ i,j,k,l,m,o \, \text{dist}} } N_{8}([[[i,j],k],l], [m,o]) \, V_{1|i,j,k,l} V_{1|m,o} + \sum_{ \substack{ 2 \leqslant i<j,k \leqslant 8 \\ 2 \leqslant i<l,m,o \leqslant 8 \\ j,k,l,m,o \, \text{dist} } } N_{8}([[i,j],k], [[l,m],o]) \, V_{1|i,j,k} V_{l,m,o} \\
        & + \sum_{ \substack{ 2 \leqslant i<j,k \leqslant 8 \\ 2 \leqslant l<m \leqslant 8 \\ 2 \leqslant l<o<q \leqslant 8 \\ i,j,k,l,m,o,q \, \text{dist} } } N_8([[i,j],k],[l,m],[o,q]) \, V_{1|i,j,k} V_{1|l,m} V_{1|o,q} \\
        & + \frac{(2\pi i)^4}{7!} \sum_{\rho \in S_7} N(1 \rho(2) \cdots \rho(8); -\frac{p_1}{2}) \, \mathcal{E}_{4}(\tau) \,, 
    \end{aligned}
    }\label{eq:8ptreg}
\end{equation}
where $\mathcal{E}_4$ is the normalised Eisenstein series, 
\begin{equation}
\label{eq:holE}
    \mathcal{E}_{2k}(\tau) := \frac{G_{2k}(\tau)}{2 \zeta(2k)} \,,
\end{equation}
which obeys $\mathcal{E}_{2k}(\tau)\to 1$ in the degeneration limit $\tau\to i\infty$.
As at lower points, all the kinematic coefficients in $\cI_8$ have been determined by matching the degeneration limit \eqref{eq:In0}. Notice that, in the last line, the numerators in the sum are evaluated at $\ell=-p_1/2$, which we will explain momentarily. This is a new feature accompanying the Eisenstein series, related to the privileged role that particle 1 takes in our worldsheet basis. The definition of \textbf{Reg} on the weight-4 functions will be presented in the next section, where we will see that it leads to the appearance of $G_{4}(\tau)$. 
More generally, $G_{2K}(\tau)$ occurs in our worldsheet basis in two ways: it appears `in its own right' as in the last line of \eqref{eq:8ptreg}, where we have its normalised version, which simplifies the numerical coefficient; and it appears as a result of \textbf{Reg}.

Let us explain the appearance, in the last kinematic coefficient of \eqref{eq:8ptreg}, of numerators evaluated at $\ell=-p_1/2$. Suppose that, instead of that last term, we wrote $C_{8,\mathcal{E}_4}\,\mathcal{E}_4$, and proceeded to extract master BCJ numerators using the degeneration limit \eqref{eq:In0}. We would find that the following combination of BCJ numerators is related to the coefficients of $\cI_8$ as
\begin{equation}\label{eq:N8ptexp}
    \frac{1}{7!} \sum_{\rho \in S_7} N(1,\rho;\ell) = N_8^{\mu\nu\rho\sigma} \left( \ell + \frac{1}{2} p_1 \right)_{\mu} \left( \ell + \frac{1}{2} p_1 \right)_{\nu} \left( \ell + \frac{1}{2} p_1 \right)_{\rho} \left( \ell + \frac{1}{2} p_1 \right)_{\sigma} + \frac{1}{(2\pi i)^4}\, C_{8,\mathcal{E}_4} \,.
\end{equation}
Evaluating this expression at $\ell=-p_1/2$, we find
\begin{equation}
    C_{8,\mathcal{E}_4} = \frac{(2\pi i)^4}{7!}\sum_{\rho \in S_7} N(1,\rho;-\frac{p_1}{2}) \,,
\end{equation}
as we wrote in the last line of \eqref{eq:8ptreg}.
The reader is not yet in a position to reproduce \eqref{eq:N8ptexp}, because we have not yet defined \textbf{Reg}, but some features are easy to anticipate. Note that
\[
\sum_{\rho\in S_7} N\big(1,\rho;\ell- \frac{1}{2} p_1\big) =
\sum_{\rho\in S_7} N\big(\rho,1;-\ell+ \frac{1}{2} p_1\big) =
\sum_{\rho\in S_7} N\big(\rho,1;\ell+ \frac{1}{2} p_1\big)\,,
\]
where the first and second equalities follow from the initial expression via the reflection \eqref{eq:Nrefl} and quasi-cyclic \eqref{eq:cyclic} properties, respectively. From the last equality, it is clear that the initial expression is an even function of $\ell$. Hence, the left-hand side in \eqref{eq:N8ptexp} must be an even function of $\ell+ \frac{1}{2} p_1$, which is indeed the case as given by the right-hand side.

We now move on to the 8-point analogues of the identity \eqref{eq:6ptidG2} at 6 points and the identities \eqref{eq:7ptidG2} at 7 points. We find\footnote{Hopefully, the meaning of the sums on the left-hand side is clear. Let us consider the last two lines: for $\rho\in S_{6;[[1,2],3}$ the sum is over permutations of the 6 elements $\{[[1,2],3],4,5,6,7,8\}$, whereas for $\rho\in S_{6;[1,2],[3,4]}$ the sum is over permutations of the 6 elements $\{[1,2],[3,4],5,6,7,8\}$.}
\begin{equation}
    \label{eq:8ptidG2}
    \boxed{
    \begin{aligned}
        & \frac{1}{8!} \sum_{\rho\in S_8} N(\rho; \ell) \big|_{\ell^{2}} =  
        \frac{1}{2} \, N_8^{\mu\nu\rho\sigma}\, 
        \ell_{\mu} \ell_{\nu} \sum_{i=1}^{8} p_{i,\rho} p_{i,\sigma} \,, \quad \forall \ell\,, \\
        & \frac{1}{7!} \sum_{\rho\in S_{7;[i,j]}} N(\rho; \ell) \big|_{\ell^{1}} = 
        \frac{1}{4} \, N_8^{\mu\nu\rho}([i,j])\,
        \ell_{\mu} \left( 2 p_{i,\nu} p_{j,\rho} + \sum_{k=1}^{8} p_{k,\nu} p_{k,\rho} \right) \,, \quad \forall \ell\,, \\
        & \frac{1}{6!} \sum_{\rho\in S_{6;[[i,j],k]}} N(\rho;\ell) \big|_{\ell^{0}} =
        \frac{1}{12} N_8^{\mu\nu}([[i,j],k])  
        \left( 2 p_{i,\mu} p_{j,\nu} + 2 p_{i,\mu} p_{k,\nu} + 2 p_{j,\mu} p_{k,\nu} + \sum_{l=1}^{8} p_{l,\mu} p_{l,\nu} \right) \,, \\
        & \frac{1}{6!} \sum_{\rho\in S_{6;[i,j],[k,l]}} N(\rho;\ell) \big|_{\ell^{0}}  =
        \frac{1}{12} N_8^{\mu\nu}([i,j],[k,l]) 
        \left( 2 p_{i,\mu} p_{j,\nu} + 2 p_{k,\mu} p_{l,\nu} +\sum_{m=1}^{8} p_{m,\mu} p_{m,\nu} \right) \,.
    \end{aligned}
    }
\end{equation}
These identities can be understood as the vanishing, required by modularity, of the kinematic coefficients $C^{\cdots}_{8,G_2\cdots}$ if we consider adding the following term to $\cI_8$:
\begin{equation}
    \begin{aligned}
      G_2(\tau) \Bigg(  & C_{8,G_2LL}^{\mu\nu} \, \reg\left\lbrack L_{(\mu} L_{\nu)} \right\rbrack - 
        \sum_{2 \leqslant i<j \leqslant 6} C_{8,G_2LV_{1|i,j}}^\mu \, \reg \left\lbrack L_{\mu} \, V_{1|i,j} \right\rbrack \\
        & + \sum_{ \substack{2 \leq i < j , k \leq 6 \\  j \neq k } } C_{8,G_2V_{1|i,j,k}}\, V_{1|i,j,k} + \sum_{ \substack{ 2 \leqslant i<j \leqslant 6 \\ 2 \leqslant i<k<l \leqslant 6 \\ j \neq k,l } } C_{8,G_2 V_{1|i,j} V_{1|k,l}} \, V_{1|i,j} V_{1|k,l} \Bigg) \,. 
    \end{aligned} 
\end{equation}


\section{$n$-point chiral integrand}
\label{sec:npt}

In this section, we will describe the $n$-point chiral integrand $\cI_n$. This will require that we present a full description of our worldsheet basis. One question is how to obtain `regularised' basis elements without closed cycles, and we will give a general prescription for this. Another question is whether  modular cusp forms should be included in the worldsheet basis. These may appear first at 16 points. We do not know the answer, but we will determine in terms of field-theory BCJ numerators all the kinematic coefficients of the chiral integrand in our worldsheet basis that are not associated to cusp forms.

\subsection{Worldsheet basis elements}
\label{subsec:basis}

At $n$ points, the basis elements take the form
\[
\label{eq:basis}
\boxed{\,
\gamma_{2K}(\tau)\,\reg\left[
L_{\mu_1}L_{\mu_2}\cdots L_{\mu_{w_L}}
\prod_{r=1}^v V_{1|i_1^{(r)}i_2^{(r)}\cdots i_{s_r}^{(r)}}
\right]\,,
\,}
\]
where both $L_\mu$ and $V_{1|ij\cdots k}$ were defined in section~\ref{sec:elliptic}. The definition of $\reg$ will be presented in section~\ref{subsec:reg}. The object $\gamma_{2K}(\tau)$ in front of the expression above is any element of weight $2K$ of the following basis of holomorphic modular forms.
\[
\label{eq:tablemodular}
\begin{tabular}{|c|c|c|c|}
\hline
weight & modular forms \\
\hline
0 & 1 \\
\hline
2 & none \\
\hline
$4\leq 2K < 16$ & ${\mathcal E}_{2K}(\tau)$ \\
\hline
$16\leq 2K$ & $\{{\mathcal E}_{2K}(\tau),\,\text{weight-2$K$ cusp forms?}\}$ \\
\hline
\end{tabular}
\]
The normalised holomorphic Eisenstein series ${\mathcal E}_{2K}(\tau)$ was defined in \eqref{eq:holE}, and both $L_\mu$ and $V_{1|ij\cdots k}$ were defined in section~\ref{sec:elliptic}. Starting at 16 points, we may in principle also admit cusp forms, which vanish as $\tau\to i\infty$. We leave that discussion to section~\ref{subsec:cusp}.

In order to be admissible as a basis element at $n$ points, the object \eqref{eq:basis} must obey the following two conditions.
\begin{enumerate}
    \item The basis element must have weight $n-4$:
\[
2K + w_L + \sum^v_{r=1}(s_r-1) = n-4\,.
\]
Including the weight-0 cases ($V_{1|i}=1$) in the product of $v$ instances of $V$'s, the indices $i_\text{a}^{(r)}$ are all distinct and, together with 1, make up the $n$ punctures, so that $1+\sum^v_{r=1} s_r=n$. The total weight condition above then leads to
\[\label{eq:kwsrelation}
v= 3+ 2K + w_L\,.
\]
\item Within each $V$ in the product, the first index is the smallest, that is, $1<i_1^{(r)}<i_\text{a}^{(r)}, \forall \text{a}>1$\,. This restriction is a consequence of Fay relations among the $V$'s just like we mentioned in section~\ref{sec:elliptic}.
\end{enumerate}
The admissible basis elements have already been exemplified in the expressions for the chiral integrand from 4 to 8 points in previous sections.\footnote{For illustration, at 5 points, we must have $K=0$ because $v\leq n-1$. The basis elements with $w_L=0$ give a product of $V$'s that equates $V_{1|i,j}$ for $1<i<j$, because $V_{1|i}=1$; this was already exemplified in section~\ref{sec:elliptic}. The basis element with $w_L=1$ is simply $\reg[L_\mu]=L_\mu$, because the product of $V$'s is $1=\prod_{i=2}^5 V_{1|i}$.}

\subsection{Kinematic coefficients for non-cusp basis elements}

We consider here the basis elements \eqref{eq:basis} where the modular form $\gamma_{2K}(\tau)$ is not a cusp form.

We start with weight 0, for which we chose to set $\gamma_{2K}(\tau)=1$. The corresponding kinematic coefficients in the chiral integrand $\cI_n$ are
\[
\label{eq:NK0}
\boxed{
(-1)^{n-w_L} N_n^{\mu_1\mu_2\cdots \mu_{w_L}} (I_1,I_2,\cdots,I_v)
\,, \quad \text{for} \; K=0 \,,}
\]
where
\[
I_r := [[\cdots[[[i_1^{(r)},i_2^{(r)}],i_3^{(r)}],i_4^{(r)}],\cdots],i_{s_r}^{(r)}]
\]
is associated to $V_{1|i_1^{(r)}i_2^{(r)}i_3^{(r)}i_4^{(r)}\cdots i_{s_r}^{(r)}}$. Just like the order of the $V$'s in \eqref{eq:basis} is irrelevant, so is the order of the $I$'s in \eqref{eq:NK0}. We define
\[
\label{eq:defNK0}
N_n^{\mu_1\mu_2\cdots \mu_{w_L}} (I_1,I_2,\cdots,I_v) :=  \frac1{w_L!} \frac{\partial^{w_L}}{\partial\ell_{\mu_1}\partial\ell_{\mu_1}\cdots \partial\ell_{\mu_{w_L}}} N(1,I_1,I_2,\cdots,I_v;\ell)
\,,
\]
which is guaranteed to be $\ell$-independent from the properties of the BCJ numerators. In words, \eqref{eq:defNK0} is the coefficient of the leading piece in $\ell$ of the BCJ numerator of the $(1+v)$-gon diagram whose corners are the particle 1 and the $v$ trees with trivalent structure $I_r$; the order of these corners is irrelevant for the leading-in-$\ell$ piece. The sign in front of \eqref{eq:NK0} can also be understood as the total number of commutators in the $I_r$'s. We have provided plenty of examples of such kinematic coefficients up to 8 points in previous sections. In those examples of \eqref{eq:NK0}, we omitted any $I_r$ with $s_r=1$ for notational simplicity.\footnote{For instance, at 5 points in \eqref{eq:I5ansatz}, we recall that we wrote $-N([2,3])$ instead of $-N(1[2,3]45)$ for a coefficient. In fact, the ordering of the four elements $\{1,[2,3],4,5\}$ is irrelevant there.}

We mentioned that \eqref{eq:NK0} is independent of the ordering of the $I_r$, because it comes from the leading-in-$\ell$ piece of a BCJ numerator. For the kinematic coefficients associated to objects \eqref{eq:basis} where $\gamma_{2K}$ is the normalised holomorphic Eisenstein series $\mathcal{E}_{2K}$ of weight $2K \geqslant 4$, this permutation invariance must also hold, again mirroring the fact that the order of the $V$'s in the basis element is irrelevant. However, for a fixed weight and set of $I_r$, the basis elements with $\gamma_{2K}=\mathcal{E}_{2K}$ have fewer powers of $L^{\mu}$ than their $\gamma_{2K}=1$ counterparts; hence, the associated kinematic coefficients originate from subleading-in-$\ell$ pieces of the numerators. These subleading pieces generally depend on the ordering of the $I_r$, so the permutation invariance must be enforced explicitly. The kinematic coefficients, which highlight the special role of particle 1 in our basis, are
\[
\label{eq:NG2K}
\boxed{
(-1)^{n-w_L} \frac{(2\pi i)^{2K}}{v!} \sum_{\rho\in S_{v;\,I_1,I_2,\cdots,I_v}} N_n^{\mu_1\mu_2\cdots \mu_{w_L}} (\rho;-\frac{p_1}{2})
\,, \quad \text{for} \; \gamma_{2K}(\tau)={\mathcal E}_{2K}(\tau) \,, }
\]
where 
\[
N_n^{\mu_1\mu_2\cdots \mu_{w_L}} (I_1,I_2,\cdots,I_v;-\frac{p_1}{2}) :=  \frac1{w_L!} \frac{\partial^{w_L}}{\partial\ell_{\mu_1}\partial\ell_{\mu_1}\cdots \partial\ell_{\mu_{w_L}}} N(1,I_1,I_2,\cdots,I_v;\ell)\Bigg|_{\ell=-\frac{p_1}{2}}
\,.
\]
We have seen the simplest example of such a kinematic coefficient at 8 points in the last line of \eqref{eq:8ptreg}. In that section, we explained the appearance of BCJ numerators evaluated at $\ell=-\frac{p_1}{2}$, and this feature extends to higher points in \eqref{eq:NG2K}. For each kinematic coefficient accompanying a basis element with $\mathcal{E}_{2K}$, there is an $\ell$-dependent permutation sum that is restricted --- due to the reflection and quasi-cyclic properties of the numerators --- to be a function of $(\ell + \frac{p_1}{2})$, and the evaluation of this permutation sum at $\ell=-\frac{p_1}{2}$ determines the kinematic coefficient.\footnote{As another example of \eqref{eq:NG2K}, $\cI_9$ includes the following terms:
$$
(2\pi i)^{4}\,\mathcal{E}_{4}(\tau) \left(\frac1{8!}\sum_{\rho\in S_{8}} N_9^\mu(\rho,-\frac{p_1}{2})\,L_\mu - \sum_{2 \leqslant i<j \leqslant 7}\frac1{7!} \sum_{\rho\in S_{7;[i,j]}} N_9(\rho,-\frac{p_1}{2})  \, V_{1|ij} \right)\,.
$$
}
Let us make one more remark about the kinematic coefficients \eqref{eq:NG2K}.
In the next subsection, we will present the prescription for the regularisation $\reg$, which involves the holomorphic Eisenstein series. In appendix~\ref{app:altreg}, we will discuss an alternative regularisation that includes also the non-modular form $G_{2}(\tau)$. As shown there, with that choice of regularisation we have examples of \eqref{eq:NG2K} starting already at 6 points.

Notice that we can consider the case \eqref{eq:NK0} as being included in a natural extension of \eqref{eq:NG2K} to the weight-0 modular form $\gamma_0(\tau)=1$.\footnote{Indeed, extending the definition \eqref{eq:G2k} of $G_{2K}(\tau)$ to $K=0$ via $G_{0}(\tau):=-g^{(0)}(0,\tau)=-1$, we obtain ${\cal E}_0=G_{0}/(2\zeta(0))=1$.} The permutation sum in \eqref{eq:NG2K} is unnecessary for $K=0 \Leftrightarrow v=3+w_L$, because the leading-in-$\ell$ piece of the numerators is already permutation invariant; and the evaluation at $\ell=-\frac{p_1}{2}$ is also no longer needed, because the tensor $N_n^{\mu_1\mu_2\cdots \mu_{w_L}} (I_1,I_2,\cdots,I_v)$ appearing in \eqref{eq:NK0} is independent of $\ell$.

We have achieved our goal of fixing in terms of field-theory BCJ numerators all the kinematic coefficients of the chiral integrand that are not associated to a cusp form.

\subsection{Regularisation of closed cycles}
\label{subsec:reg}

We will finally present the general prescription of the regularisation $\reg$, on which our worldsheet basis relies.

To begin with, note that closed cycles never appear in basis elements involving only $V$-functions. In a basis element \eqref{eq:basis}, every closed cycle arises from multiplying $\egg{1,i}$ in $L$ with $V$-functions, or with other $L$'s. The essential part of \reg{} is to shift the $\egg{1,i}$'s in $L^{\mu}$ by a derivative-like operator,
\begin{equation}
\label{eq:Lhat}
    \boxed{ \,
    \hat{L}_{\mu} := 2\pi i \, \ell_{\mu} + \sum_{i=2}^{n} p_{i,\mu} \left( \egg{1,i} - \hat{D}_i \right) \, 
    } \,,
\end{equation}
such that
\begin{equation}
\label{eq:naivereg}
        \reg^{\text{(naive)}} \lbrack L_{\mu_1} \cdots L_{\mu_{w_L}} \prod_{S} V_{1|S} \rbrack 
        := \hat{L}_{\mu_1} \cdots \hat{L}_{\mu_{w_L}} \prod_{S} V_{1|S} \,,
\end{equation}
where the `naive' qualification will be explained later.
We take $\hat{D}_i$ to be a linear operator acting as $\partial_{z_i}$ on $\eg{m}{1,i}$,
with the explicit definition to be given in \eqref{eq:fulldhat}. The idea is that, each time we act with $\hat{L}_{\mu}$, the function being acted upon is already regularised, i.e.~free of closed cycles. Therefore, it suffices to define $\hat{D}_i$ on monomials of $g$-functions that do not contain closed cycles. 

To motivate the definition of $\hat{D}_i$, we first consider a simple example, namely when a $\egg{1,i}$ in $\hat{L}_{\mu}$ is involved in a two-cycle: $\egg{1,i}\eg{m}{1,i}$, with $m\geq1$. The identity
\begin{equation}
\label{eq:derivativenew}
    \left( \egg{1,i} - \partial_{i} \right) \eg{m}{1,i} = (m+1) \eg{m+1}{1,i} - \sum_{ \substack{k=2 \\ k \text{ even}} }^{m+1} G_k(\tau) \eg{m+1-k}{1,i} \,,
\end{equation}
where the right-hand side is free of closed cycles, suggests, together with \eqref{eq:Lhat}, that we define $\hat{D}_i \eg{m}{1,i} := \partial_{i} \eg{m}{1,i}$. Note, however, that we cannot simply take $\hat{D}_i$ to be $\partial_{z_i}$, as the latter would act on other $g$-functions involving the puncture $i$, i.e.~$\eg{m}{i,j}$ with $j \neq 1,i$. This would introduce extra two-cycles because
\begin{equation}
    \partial_i \eg{m}{i,j} = -\egg{i,j}\eg{m}{i,j} + (m+1) \eg{m+1}{i,j} - \sum_{ \substack{k=2 \\ k \text{ even}} }^{m+1} G_k(\tau) \eg{m+1-k}{i,j} \,.
\end{equation}
A better choice for $\hat{D}_i$ would be
\begin{equation}\label{eq:naivedhat}
    \sum_{m=1}^{\infty} \frac{\partial \eg{m}{1,i}}{\partial z_i} \frac{\partial}{\partial \eg{m}{1,i}} \,,
\end{equation}
but this is still naive, because a chain of the form
\begin{equation}
    \eg{m_0}{1,i_1} \eg{m_1}{i_1,i_2} \cdots \eg{m_k}{i_{k},i}
\end{equation}
also involves $\eg{\bullet}{1,i}$ implicitly through Fay identities:
\begin{align}\label{eq:faygeneral}
        \eg{m}{1,j} \eg{m'}{i,j} =
       & - (-1)^{m'} \eg{m+m'}{1,i} 
         + \sum_{a=0}^{m} (-1)^{a} \binom{m'+a-1}{a} \eg{m-a}{1,i} \eg{m'+a}{i,j} \nonumber\\
        &\qquad  + \sum_{a=0}^{m'} (-1)^{m'} \binom{m+a-1}{a} \eg{m+a}{1,j} \eg{m'-a}{1,i} \,.
\end{align}
One can understand the Fay identities as an elliptic version of partial-fraction decomposition. Using \eqref{eq:faygeneral}, any chain can be reduced to a basis where one instance of $\eg{\bullet}{1,i}$ is manifest:
\begin{equation}\label{eq:gkkbasis}
    \begin{aligned}
        &\, \eg{m_0}{1,i_1} \eg{m_1}{i_1,i_2} \cdots \eg{m_k}{i_{k},i} \\
        =& \sum_{ \substack{ a_p \in \{ -m_p \} \cup \mathbb{Z}^{\geqslant 0} \\ \sum a_p \leqslant 0 \\ a_p \text{ not all zero } } } 
        (-1)^{1+\sum_p a_p} \prod_{p=0}^{m} \binom{m_p+a_p-1}{a_p} 
        \eg{-a_0 - \cdots - a_k}{1,i} \eg{m_0+a_0}{1,i_1} \eg{m_1+a_1}{i_1,i_2} \cdots \eg{m_k+a_k}{i_{k},i} \,.
    \end{aligned}
\end{equation}
The second line above may seem to contain a closed cycle of the form $\eg{\bullet}{1,i} \eg{\bullet}{1,i_1} \eg{\bullet}{i_1,i_2} \cdots \eg{\bullet}{i_k,i}$, but this is in fact excluded by the summation rules.
In each term, there is at least one $a_p = -m_p$, making the corresponding function $\eg{m_p+a_p}{i_{p},i_{p+1}}$ reduce to $\eg{0}{i_{p},i_{p+1}} = 1$. Therefore, each term in the second line is free of closed cycles. Furthermore, when $a_p = -m_p$, the binomial coefficient evaluates, by the standard negative-index convention, to $\binom{m_p+a_p-1}{a_p} = (-1)^{m_p-1}$.

The full definition of $\hat{D}_i$ is then
\begin{equation}\label{eq:fulldhat}
    \boxed{
    \hat{D}_i (f) := \sum_{m=1}^{\infty} \frac{\partial \eg{m}{1,i}}{\partial z_i} \frac{\partial}{\partial \eg{m}{1,i}} \left( f\big|_{\text{replacement by \eqref{eq:gkkbasis}}} \right) \,,
    }
\end{equation}
for any monomial $f$ without closed cycles. Effectively, $\hat{D}_i(f)$ exchanges the single instance of $\eg{m}{1,i}$ in each term of $f$ obtained from a 1-$i$ chain (after appropriate use of Fay relations) with $\partial_i\eg{m}{1,i}$. Since $f$ is monodromy-invariant and the monodromy property of $\partial_i\eg{m}{1,i}$ mirrors that of $\eg{m}{1,i}$, the final object is also monodromy-invariant.

We have now described all the ingredients in \eqref{eq:naivereg}, but as the `naive' qualification indicates, we are not done yet. Two challenges remain.
\begin{itemize}
    \item Firstly, as shown in \eqref{eq:derivativenew}, the derivatives of $g$-functions give rise to the quasi-modular form $G_2(\tau)$, which is not a modular form. However, we cannot drop all instances of $G_2(\tau)$ arising in this manner, because this object has a non-trivial degeneration limit, namely $G_2(\tau)\to 2\zeta(2)$  as $\tau\to i\infty$. 
    \item Secondly, as it stands, the expression on the right-hand side of \eqref{eq:naivereg} is not symmetric in the spacetime indices of the $L$'s. That is, generically, $[\hat L_\mu,\hat L_\nu]\neq0$ acting on products of $g$-functions. This is undesirable, because we are regulating an expression that is symmetric in these indices, i.e.~the argument of the left-hand side of \eqref{eq:naivereg}.
\end{itemize}
It turns out that {\it both} of these challenges can be solved with a simple prescription.\footnote{We checked explicitly that the prescription below effectively leads to $[\hat L_\mu,\hat L_\nu]$ dropping out of expressions, up to weight 18. This is not necessarily the only prescription that does the job, however.} The action of \reg{} includes a projection ${\mathcal P}_G$ of products of $G_{2K_s}(\tau)$ with $K_s\geq1$ into the single $G_{2K}(\tau)$ with the same weight:
\begin{equation}
\label{eq:PG}
         {\mathcal P}_G\Big[\prod_sG_{2K_s}(\tau)\Big]:=\frac{\prod_s\big(2\,\zeta(2K_s)\big)}{2\,\zeta(2\sum_sK_s)} \;G_{2\sum_sK_s}(\tau)\,.
\end{equation}
We take ${\mathcal P}_G[1]:=1$. After this projection, we will drop terms that include $G_2(\tau)$, due to the requirement of modularity.  
As illustrated in sections~\ref{sec:6pt} to \ref{sec:8pt}, the fact that $G_{2}(\tau)$ is ultimately absent from the superstring correlator implies non-trivial identities on field-theory BCJ numerators, which we will discuss again in section~\ref{sec:G2}.
Notice that ${\mathcal P}_G$ does not project into the cusp space; we will discuss cusp forms in the next section.

Putting all the pieces together, we define the basis elements of the chiral integrand such that
\begin{equation}
\label{eq:finalreg}
    \boxed{
        \reg \lbrack L_{\mu_1} \cdots L_{\mu_m} \prod_{S} V_{1|S} \rbrack 
        := {\mathcal P}_G \lbrack \hat{L}_{\mu_1} \cdots \hat{L}_{\mu_m} \prod_{S} V_{1|S} \rbrack \bigg|_{G_2(\tau) \mapsto 0}  
    } \,.
\end{equation}

Let us illustrate the definition of \reg{} with the examples at 6 points. There are two types of basis elements to be regularised, $\reg \left\lbrack L_{\mu} L_{\nu} \right\rbrack$ and $\reg\left\lbrack L_{\mu} V_{1|i,j} \right\rbrack$. We have
\begin{align}
        \reg & \left\lbrack L_{\mu} L_{\nu} \right\rbrack 
        = \hat{L}_{\mu} \hat{L}_{\nu} \bigg|_{G_2 \mapsto 0} = \hat{L}_{\mu} L_{\nu} \bigg|_{G_2 \mapsto 0} \nonumber\\
        &= (2\pi i)^2 \ell_{\mu} \ell_{\nu} + 4\pi i\, \ell_{(\mu} \sum_{i=2}^{6} p_{i,\nu)} \egg{1,i} + \sum_{ \substack{ i \neq j } } p_{i,\mu} p_{j,\nu} \, \egg{1,i} \egg{1,j} +  \sum_{i=2}^{6} p_{i,\mu} p_{i,\nu} \left( (\egg{1,i})^2 - \partial_i \egg{1,i} \right) \bigg|_{G_2\mapsto 0} \nonumber\\
        &= (2\pi i)^2 \ell_{\mu} \ell_{\nu} + 4\pi i\, \ell_{(\mu} \sum_{i=2}^{6} p_{i,\nu)} \egg{1,i} + \sum_{ \substack{ i \neq j } } p_{i,\mu} p_{j,\nu} \egg{1,i} \egg{1,j} +  \sum_{i=2}^{6} 2\, p_{i,\mu} p_{i,\nu} \, \eg{2}{1,i} \,,
\end{align}
where we have omitted the operator $\hat{D}_i$ in the rightmost $\hat{L}$ because there is nothing for it to act on, and 
\begin{align}
         \reg & \left\lbrack L_{\mu} V_{1|i,j} \right\rbrack = \hat{L}_{\mu} V_{1|i,j} \bigg|_{G_2 \mapsto 0} \nonumber \\
        &= 2\pi i\, \ell_{\mu} V_{1|i,j} + \sum_{k\neq i,j} p_{k,\mu} \egg{1,k} V_{1|i,j} + p_{i,\mu} \left( (\egg{1,i})^2 - \partial_{i} \egg{1,i} \right) - p_{j,\mu} \left( (\egg{1,j})^2 - \partial_{j} \egg{1,j} \right) \bigg|_{G_2\mapsto 0} \nonumber\\
        &= 2\pi i\, \ell_{\mu} V_{1|i,j} + \sum_{k\neq i,j} p_{k,\mu} \egg{1,k} V_{1|i,j} + 2\, p_{i,\mu} \, \eg{2}{1,i} - 2\, p_{j,\mu}\, \eg{2}{1,j} \,.
\end{align}
Both cases effectively correspond to  \eqref{eq:defreg6pt}.

To summarise, we constructed \reg{} with three goals: (1) to exclude closed cycles; (2) to maintain monodromy invariance; and (3) to ensure the nicest possible degeneration limit of the final basis elements. The last condition is quite implicit, so let us explain it with a simple example. In the 8-point chiral integrand \eqref{eq:8ptreg}, which has weight 4, $G_4(\tau)$ arises in two manners: in `its own right', in the last term, where we normalised it to simplify the kinematic coefficient; and within basis elements obtained with \reg{}, due to \eqref{eq:derivativenew}. The latter appearance is a choice, because monodromy invariance of the basis elements would still hold had we dropped $G_4(\tau)$ contributions arising in \reg{}. This would, however, change the degeneration limit.
The `nice' degeneration limit then means that we obtain the simplest possible kinematic coefficient for the last term in \eqref{eq:8ptreg}. Otherwise, we would have to solve a large linear system of equations to find the kinematical coefficients in terms of pieces of BCJ numerators, as we did in our previous work~\cite{Geyer:2024oeu}, which would be vastly more difficult at higher multiplicity.

As an example, the simplest case where a modular form $G_4$ arises from the regularisation \reg{} is:
\begin{align}
         \reg\left\lbrack L_{\mu_1} L_{\mu_2} L_{\mu_3} L_{\mu_4} \right\rbrack
        := &\; \mathcal{P}_{G} \left\lbrack \hat{L}_{\mu_1} \hat{L}_{\mu_2} \hat{L}_{\mu_3} \hat{L}_{\mu_4} \right\rbrack \Bigg|_{G_2(\tau)\mapsto0} \nonumber\\
        =&\; \mathcal{P}_{G} \left\lbrack \left( 2\pi i\, \ell + \sum_{i=2}^{n} p_i ( \egg{1,i} - \partial_i ) \right)_{\mu_1\mu_2\mu_3\mu_4} \right\rbrack \Bigg|_{G_2(\tau)\mapsto0} \nonumber\\
        =&\; \left( 2\pi i\, \ell + \sum_{i=2}^{n} p_i\, \egg{1,i} \right)_{\mu_1\mu_2\mu_3\mu_4} \bigg|_{(\egg{a,b})^m \mapsto m!\, \eg{m}{a,b}} \nonumber\\
        & + \left( \sum_{i=2}^{n} 9\, p_{i,\mu_1} p_{i,\mu_2} p_{i,\mu_3} p_{i,\mu_4} + \sum_{2 \leqslant i < j \leqslant n} 60\, p_{i,(\mu_1} p_{i,\mu_2} p_{j,\mu_3} p_{j,\mu_4)} \right) G_4(\tau) \,, 
\end{align}
where in the second and third lines we use the shorthand $(\cdot)_{\mu_1\mu_2\mu_3\mu_4} := (\cdot)_{\mu_1}(\cdot)_{\mu_2}(\cdot)_{\mu_3}(\cdot)_{\mu_4}$.

We present in appendix~\ref{app:proof} the sketch of a proof that our construction provides a complete basis of puncture-dependent but monodromy-invariant worldsheet functions. The proof is recursive, that is, from the $n$-point basis, we get the $(n+1)$-point basis.

Ref.~\cite{Mafra:2018pll} made a related proposal for worldsheet functions up to quadratic order in $\ell$. Their objects match ours at low multiplicity; see their equations (6.28) and (6.29). At higher points, our basis elements incorporate the holomorphic Eisenstein series, and of course also admit any order in $\ell$.

A final comment is that we could have chosen to keep $G_2(\tau)$ in \reg{}, that is, to ignore the final command in \eqref{eq:finalreg}; we will look into this in appendix~\ref{app:altreg}. Ultimately, any contributions with $G_2(\tau)$ must vanish by modularity, but the alternative regularisation provides a different derivation of the $G_2(\tau)$ identities among BCJ numerators, which we will return to in section~\ref{sec:G2}.

\subsection{Cusp forms and the ambiguity of the field-theory limit}\label{subsec:cusp}

Here, we consider the worldsheet basis elements \eqref{eq:basis} for which $\gamma_{2K}(\tau)$ is a cusp modular form, meaning that it vanishes in the field-theory degeneration $\tau\to i\infty$.

It is easy to see that cusp forms may exist starting at weight $n-4=12$, which corresponds to 16 points. We start by recalling that holomorphic modular forms are fixed-weight polynomials of ${\mathcal E}_{4}(\tau)$ and ${\mathcal E}_{6}(\tau)$. At a given weight$\,\geq4$, the holomorphic Eisenstein series ${\mathcal E}_{2K}(\tau)$ gives us one such basis element, and in fact for $4\leq\,$weight$\,<12$ there is a single linearly independent modular form.\footnote{We have, up to weight 10: ${\mathcal E}_4$, ${\mathcal E}_6$, ${\mathcal E}_8=({\mathcal E}_4)^2$, ${\mathcal E}_{10}= {\mathcal E}_4 {\mathcal E}_6$.} Starting at weight 12, the space of modular forms is higher-dimensional (with the exception of weight 14). 

At weight 12, we have two linearly independent modular forms; e.g.~taking these to be $({\mathcal E}_4)^3$ and $({\mathcal E}_6)^2$, we can write ${\mathcal E}_{12}=\frac{441}{691} (\mathcal{E}_4)^3 + \frac{250}{691} (\mathcal{E}_6)^2$. We use here the normalised version of the holomorphic Eisenstein series, such that ${\mathcal E}_{2K}(\tau)\to1$ in the cusp limit $\tau\to i\infty$. From this, we conclude that the following modular form vanishes at the cusp, i.e.~it is a cusp form:
\[
\label{eq:Delta}
\Delta(\tau) = \frac{(2\pi)^{12}}{12^3} \left[ \left( {\mathcal E}_{4}(\tau) \right)^3 - \left( {\mathcal E}_{6}(\tau) \right)^2 \right] = \big(60\, G_4(\tau)\big)^3 -27  \big(140\, G_6(\tau)\big)^2 = (2\pi)^{12}\,q \prod_{n=1}^\infty(1-q^n)^{24} \,.
\]
We chose the normalisation here such that $\Delta(\tau)$ is the so-called modular discriminant. 

Let us choose ${\mathcal E}_{12}$ and $\Delta$ as the two independent forms at weight 12, as in table~\eqref{eq:tablemodular}. From the basis elements \eqref{eq:basis}, we can write the $K=6$ part of $\cI_{16}$ as
\[
\label{eq:C16I12}
C_{16,{\mathcal E}_{12}} \,{\mathcal E}_{12}(\tau) + C_{16,\Delta}\, \Delta(\tau).
\]
Using \eqref{eq:NG2K}, we have
\[
C_{16,{\mathcal E}_{12}} = \frac{(2\pi i)^{12}}{11!} \sum_{\rho\in S_11} N(1,\rho;-\frac{p_1}{2})\,.
\]
On the other hand, the second term in \eqref{eq:C16I12} vanishes as $\tau\to i\infty$, and therefore does not contribute to the BCJ numerators extracted via \eqref{eq:In0}. Hence, the kinematic coefficient $C_{16,\Delta}$ cannot be determined from the field-theory limit.

The space $ M_{2K}$ of classical modular forms of weight $2K$ under SL(2,${\mathbb Z}$) has dimension
\[
\text{dim}\, M_{2K} = \left\{
\begin{aligned}
& 0\,, \qquad\qquad\quad \text{if}\;\, K\not\in \mathbb{N}_{0}\,, \\
& \lfloor K/6 \rfloor+1\,, \quad \text{if}\;\, K\geq0\,,\; K\not\equiv1 \; (\text{mod}\;6)\,, \\
& \lfloor K/6 \rfloor\,, \qquad\;\,\; \text{if}\;\, K\geq0\,,\; K\equiv1 \; (\text{mod}\;6)\,.
\end{aligned}
\right.
\label{eq:dimM2k}
\]
One can span $ M_{2K}$ with all $\,{\mathcal E}_{2K'}(\tau)\,\Delta(\tau)^m\,$ such that $2K=2K'+12m$, including ${\mathcal E}_0:=1$ and excluding the non-modular form ${\mathcal E}_2(\tau)$. The latter's exclusion is responsible for the separate case of the last line in \eqref{eq:dimM2k}. For instance, $ M_{14}$ is one-dimensional, spanned only by $\mathcal{E}_{14} = \mathcal{E}_4^2 \mathcal{E}_6$. Weight 14 is the highest for which a cusp form is absent. Note, however, that at 18 points we have cusp worldsheet basis elements, given by \eqref{eq:basis} with $\Delta(\tau)$ times weight-2 regularised products of $L$'s and/or $V$'s.

We conclude that, if we consider all basis elements \eqref{eq:basis} that are in principle allowed in the superstring worldsheet correlator, the kinematic coefficients associated to $\gamma_{2K}(\tau)$ being a cusp form --- which, as mentioned, appear starting at 16 points --- cannot be  determined from the field-theory limit. It could happen that all such kinematic coefficients vanish, such that cusp forms are excluded from the correlator, but we have no indication of it at present. The difficulty in performing first-principles calculations of the worldsheet correlator at high multiplicity makes it hard to settle this question.

Given the paucity of explicit higher-point computations, it is worth briefly mentioning how classical modular forms arise in correlators, even if we will not arrive at any result here. We consider the RNS formalism for concreteness, where the modular forms can occur in two ways. The first is via the sums over the even spin-structures of the worldsheet fermions. These sums take the general form
\[
\label{eq:sumS}
{\mathscr S}(x_1,\cdots,x_m;\tau) := \sum_{\nu=1,2,3} (-1)^\nu \left(\frac{\theta_{\nu+1}(0,\tau)}{\theta'_1(0,\tau)}\right)^4\, S_\nu(x_1,\tau)\,S_\nu(x_2,\tau)\,\cdots\, S_\nu(x_m,\tau)\,,
\]
where each argument $x_r$ is identified with $z_i-z_j$ for some pair $(i,j)$ of punctures, with the restriction $\sum_{r=1}^m x_r=0$\,. The Szeg\H{o} kernels denoted as $S_\nu$ are the fermionic Green's functions for each spin structure $\nu$, and are defined as
\[
S_\nu(z,\tau):= \frac{\theta_1'(0,\tau)\,\theta_{\nu+1}(z,\tau)}{\theta_{\nu+1}(0,\tau)\, \theta_1(z,\tau)}\,.
\]
The sums \eqref{eq:sumS} have been worked out in \cite{Tsuchiya:1988va,Tsuchiya:2012nf} in terms of the Weierstrass $\wp(z,\tau)$ function and its $z$-derivatives, and of course the modular forms whose ring is generated by $G_4(\tau)$ and $G_6(\tau)$.\footnote{Note that $\wp(z,\tau)=-\partial_z^2 \log \theta_1(z,\tau) - G_2(\tau)$.} More conveniently for us, ref.~\cite{Broedel:2014vla} reworked the sums in terms of objects that directly fit our ansatze, namely the functions $g^{(w)}(z,\tau)$ and the modular forms. The crucial point is that, whatever the specific sum, at each modular weight only one polynomial of $G_4(\tau)$ and $G_6(\tau)$ contributes. For concreteness, let us quote the result of the sums \eqref{eq:sumS} as presented in \cite{Broedel:2014vla}:
\[
\label{eq:sumSdone}
{\mathscr S}(x_1,\cdots,x_m;\tau) = V_{m-4}(x_1,\cdots,x_m;\tau) + \sum_{K=2}^{\left\lfloor \frac{m}{2} \right\rfloor -2} \mathscr{G}_{2K}(\tau) \, V_{m-4-2K}(x_1,\cdots,x_m;\tau)\,,
\]
where the functions $V$ are monodromy-invariant polynomials of the $g^{(w)}_{ij}$'s, and the functions $\mathscr{G}_{2K}(\tau)$ are modular forms of weight $2K$, not identical to $G_{2K}(\tau)$. In detail, the functions $V$ can be defined in terms of the Kronecker-Eisenstein series \eqref{eq:F},
\[
\label{eq:VpFF}
V_{p}(x_1,\cdots,x_m;\tau) := \left(\,\prod_{r=1}^m F(x_r,\eta,\tau)\right)\Bigg|_{\eta^{p-m}}\,,
\]
where we pick up the coefficient of $\eta^{p-m}$ in an expansion in small $\eta$.\footnote{In \cite{Broedel:2014vla}, the functions $V$ were defined in terms of the non-holomorphic but doubly-periodic cousin of $F(z,\eta,\tau)$. As mentioned there, however, the definitions are equivalent, subject to the constraint $\sum_{r=1}^m x_r=0$.} The case $p=m-2$ coincides with the $V$'s from previous sections; recall \eqref{eq:VFF}. Regarding the modular forms $\mathscr{G}_{2K}(\tau)$, their definition can be found in \cite{Broedel:2014vla}. The lowest weight cases are: $\mathscr{G}_{4}= 3\, G_4,\;
\mathscr{G}_6 = 10 \,G_6, \;
\mathscr{G}_8 = 42\, G_8,\;
\mathscr{G}_{10} = 168 \,G_{10}, \;
\mathscr{G}_{12} = 627\, G_{12} + 9\, (G_4)^3$\,.

Along the spin-structure sums over fermionic correlators, there is another instance where modular forms arise in superstring correlators: in differential and algebraic relations satisfied by the functions $g^{(w)}(z,\tau)$ that appear in the worldsheet basis. See \eqref{eq:derivativenew} for a differential relation that includes modular forms, which we exploited to define the regularisation $\reg$. In addition, we note that the Fay identities that follow from \eqref{eq:Fay} lead to algebraic relations like
\[
g^{(2)}_{12}\, g^{(2)}_{12} - 2\, g^{(1)}_{12}\, g^{(3)}_{12} + 2\, g^{(4)}_{12} - 3 \, G_{4}(\tau) = 0\,,
\]
when we consider $z'\to z$. This necessarily involves closed cycles, so does not muddle up the appearance of $g$-functions and the Eisenstein series in our natural basis. It could, however, potentially play a role in how classical modular forms appear in first-principles (e.g.~RNS) derivations of the correlators.

It is hard to foresee the full extent of simplifications in the appearance of modular forms in actual superstring correlators. Perhaps the pure-spinor formalism, where we do not have sums over spins structures, is a more promising avenue to investigate this. Nevertheless, we have already seen some examples of simplifications that are associated to our natural worldsheet basis. For instance, the appearance of modular forms, within $\reg$ and via $\gamma_{2K}(\tau)$, conspires to produce remarkably simple kinematic coefficients associated to the field-theory limit. 
In view of this elegance, one may hope that simplifications occur for the kinematic coefficients of cusp forms. The greatest simplification of all would be that these vanish. There may be some ambiguity even here, however. We employed the projection ${\mathcal P}_G$ in \eqref{eq:PG}, into the holomorphic Eisenstein series. While the projection appears to us to be well motivated, as we discussed then, we cannot exclude that there are alternative prescriptions. These would lead to a different choice of basis, where the splitting between non-cusp and cusp basis elements would be altered. This would, therefore, raise the question of in which basis the coefficients of cusp forms would vanish.

Let us make one final comment on cusp forms. We mentioned earlier that the superstring chiral integrand can be imported into the ambitwistor string. Since the ambitwistor moduli-space integral localises on the field-theory degeneration \cite{Geyer:2015bja,Geyer:2015jch,Geyer:2016wjx,Geyer:2018xwu}, it is insensitive to cusp forms, so it says nothing about whether these may be present in the superstring or not. We see immediately that importing an ambitwistor-string chiral integrand into the conventional superstring requires more care, due to the possible ambiguity of cusp forms.


\section{$G_2(\tau)$ identities conjecture at $n$ points}
\label{sec:G2}

As noted in the previous sections, the modularity of the chiral integrand imposes additional constraints on the BCJ numerators of maximal super-Yang-Mills/supergravity, beyond the standard Jacobi relations and automorphic properties (reflection and quasi-cyclicity). These extra relations arise from the requirement that the quasi-modular form $G_2(\tau)$ drops out of the superstring correlator. More explicitly, at $n$ points, we require that the kinematic coefficients of the following `would-be' basis elements vanish:
\[
\label{eq:G2basis}
\boxed{\,
G_{2}(\tau)\,\reg\left[
L_{\mu_1}L_{\mu_2}\cdots L_{\mu_{w_L}}
\prod_{r=1}^v V_{1|i_1^{(r)}i_2^{(r)}\cdots i_{s_r}^{(r)}}
\right]\,,
\,}
\]
with, according to \eqref{eq:kwsrelation},
\begin{equation}
    v = 5+w_L \,.
\end{equation}
The kinematic coefficient of each basis element above is related to the fully symmetrised BCJ numerators of the $(v+1)$-gon trivalent diagram. Taking this `would-be' coefficient to vanish produces the following identity:
\begin{equation}\label{eq:generalG2id}
    \begin{aligned}
       & \frac{1}{(v+1)!} \sum_{\rho \in S_{1+v};\,I_1,\cdots,I_v,I_{v+1}} N(\rho;\ell)\big|_{\ell^{v-5}} \\
        & \quad  = \,  \frac{1}{12} \binom{v-3}{2} N(I_1,\cdots,I_v,I_{v+1};\ell)\big|_{ \ell_{\mu_1} \cdots \ell_{\mu_{v-3}} }\;  \ell_{\mu_3} \cdots \ell_{\mu_{v-3}} \left( \sum_{r=1}^{v+1} p_{I_r,\mu_1} p_{I_r,\mu_2} \right) \,,
    \end{aligned}
\end{equation}
where every $I_r$ corresponds to a (massless or massive) corner, and $p_{I_r} := \sum_{i \in I_r} p_{i}$ denotes the total momentum entering each corner. To be precise, requiring the coefficients of \eqref{eq:G2basis} to vanish produces \eqref{eq:generalG2id} with $I_1 = \{1\}$. For other cases where particle 1 is not single out, the corresponding identities can be obtained from this special case via the reflection \eqref{eq:Nrefl} and quasi-cyclic \eqref{eq:cyclic} properties.

Recall that the BCJ numerator of a $(v+1)$-gon is a polynomial of order $v-3$ in the loop momentum $\ell$.
Therefore, \eqref{eq:generalG2id} is relating the subsubleading $\ell$ part of the BCJ numerators and the leading $\ell$ part. In particular, the subsubleading $\ell$ part of the permutation sum over the corners of the $(v+1)$-gons is determined from the leading $\ell$ part, which is independent of the ordering of the corners. 

This provides a straightforward $n$-point extension of the `$G_2(\tau)$ identities' we found up to 8 points --- in particular, in \eqref{eq:6ptidG2} at 6 points, in \eqref{eq:7ptidG2} at 7 points, and in \eqref{eq:8ptidG2} at 8 points. It would be interesting to understand the meaning of these field-theory relations independently of their origin in the modularity of the superstring.


\section{Conclusion}
\label{sec:conclusion}

Let us summarise the results of this paper, where we studied one-loop supertring amplitudes for massless external states.
\begin{itemize}
    \item We constructed a worldsheet basis for $n$-point superstring correlators at genus one in the chiral-splitting representation. This provides the one-loop generalisation of the `Parke-Taylor basis' for the tree-level amplitude \eqref{eq:Atree}. Our basis builds on earlier results, especially the `generalised elliptic' approach of Mafra and Schlotterer \cite{Mafra:2017ioj,Mafra:2018pll}, extended here to $n$ points. In particular, we described how to systematically include the loop momentum at any multiplicity, and how to incorporate in a natural manner the holomorphic Eisenstein series.
    \item We identified the kinematic coefficients of our genus-one non-cusp basis elements with pieces of one-loop BCJ numerators of the loop-integrand-level field-theory limit. The existence of one-loop BCJ numerators in the field theory follows from the general structure of the worldsheet correlator. This result provides the one-loop generalisation of the kinematic numerators in the Parke-Taylor basis of the tree-level amplitude \eqref{eq:Atree}.
    \item We discussed why we could not fix the kinematic coefficients of cusp basis elements, which vanish in the field-theory degeneration, and which are in principle admissible starting at 16 points. Whether or not these coefficients vanish in the superstring correlator answers the question of whether or not the superstring amplitude can be thought of as a `trivial' $\alpha'$ dressing of field theory at $n$ points. That is, whether the knowledge of the field-theory loop integrand completely determines the superstring moduli-space integrand. 
    \item We showed how the modularity of the superstring amplitude imposes constraints on the field-theory loop integrand of maximal super-Yang-Mills and supergravity (also in fewer than 10 spacetime dimensions by dimensional reduction). In particular, the non-modular cousin $G_2(\tau)$ of the Eisenstein series, $G_{2K}(\tau)$ with $K\geq2$, must be absent from the superstring correlator. As a result, the one-loop BCJ numerators obtained in the field-theory degeneration are more constrained than what may be naively expected from a purely field-theory perspective. It would be good to understand the physical interpretation of these constraints in field theory.
\end{itemize}
These results achieve the goal we started pursuing in \cite{Geyer:2024oeu} of determining the one-loop superstring correlator in terms of the field-theory limit. That project was itself a one-loop higher-point counterpart to the three-loop four-point result of~\cite{Geyer:2021oox}.

There are various possible directions for the future. Our regularisation of the genus-one basis elements may admit a more elegant definition that exhibits manifestly the property of monodromy invariance; previous work on generalised elliptic integrands may hold clues \cite{Mafra:2018pll,Mafra:2022unp}. Still at one loop, while we described the structure of the superstring correlator, including of the kinematic coefficients that are pieces of BCJ numerators, we have not constructed $n$-point explicit expressions for the BCJ numerators; the state-of-the-art at one loop is ref.~\cite{Edison:2022jln}. We expect, however, that our decomposition of the numerators and the new constraints we presented will both be useful. In addition, at one loop, we can explore the connection of our results to recent work on KLT relations and twisted cohomology \cite{Mizera:2017cqs,Mizera:2019blq,Casali:2019ihm,Stieberger:2022lss,Stieberger:2023nol,Bhardwaj:2023vvm,Mazloumi:2024wys} and on new ways of organising loop integrands \cite{Bern:2024vqs,Arkani-Hamed:2024tzl,Cao:2024olg,Cao:2025ygu,Xie:2025utp}. One may also study the amplitude --- i.e.~after moduli-space integration --- in the small $\alpha'$ expansion \cite{Mafra:2019ddf,Mafra:2019xms,Broedel:2018izr,Gerken:2020yii} or even at finite $\alpha'$ \cite{Eberhardt:2023xck,Baccianti:2025whd}. The two-loop problem is a natural future step, and there is significant recent work on the superstring correlator \cite{DHoker:2020prr,DHoker:2020tcq,DHoker:2021kks,DHoker:2022xxg,DHoker:2023khh,DHoker:2025jgb}.
At three loops, questions remain about the chiral measure as recently discussed in \cite{Dunin-Barkowski:2025rda}, which revisited a point raised in \cite{Witten:2015hwa}. Nevertheless, a relatively simple measure proposed in \cite{Cacciatori:2008ay} formed the basis for the three-loop conjecture of \cite{Geyer:2021oox} matching the known field-theory limit; this measure may be the end result of significant simplications when starting from first principles. Moreover, there is the issue of non-projectability of supermoduli space at higher genus \cite{Donagi:2013dua}, which presents an obstacle to the chiral-splitting framework, but which may also provide a path to its generalisation. 

Beyond these directions, there is an overall motivation in this line of work, namely that the remarkably close connection of superstring theory to its field-theory limit may be one of its determining features.

\subsection*{Acknowledgements}

We are grateful to Yvonne Geyer and Jiachen Guo for their collaboration in ref.~\cite{Geyer:2024oeu}, and also to Oliver Schlotterer for comments on the draft and for many discussions that helped our understanding of this topic. 
RM and LR are supported by the Royal Society, via a University Research Fellowship and a Newton International Fellowship, respectively. This work is also supported by the UK's Science and Technology Facilities Council (STFC) Consolidated Grants ST/T000686/1 and ST/X00063X/1 ``Amplitudes, Strings \& Duality".

\appendix


\section{Comment on choosing $I_n$ to be independent of $\alpha'$}
\label{app:Ialpha}

We have claimed that it is possible to express the one-loop superstring worldsheet correlator in the chiral-splitting representation \eqref{eq:Ac}-\eqref{eq:Ao} such that a chiral integrand $\cI_n$ is independent of $\alpha'$. That is, the $\alpha'$ dependence is carried solely by the Koba-Nielsen factor and the normalisation of the amplitude. The argument is that $\cI_n$ is a polynomial in $1/\alpha'$ that should be finite as $\alpha'\to0$, up to total derivatives, such that there is a finite integrand-level field-theory limit that reproduces the loop integrand in super-Yang-Mills/supergravity.

Here, we discuss this from the point of view of a practical calculation, at least schematically. For concreteness, let us consider the amplitude for the scattering of NS states:
\[
V\sim \Big( \epsilon\cdot \partial X + \frac{\alpha'}{2} \, k\cdot \Psi \, \epsilon\cdot\Psi \Big) e^{ik\cdot X} = \alpha' \Big( \epsilon\cdot \partial (X/\alpha') + \frac1{2} \, k\cdot \Psi \, \epsilon\cdot\Psi \Big) e^{ik\cdot X}\,.
\]
The only OPE of worldsheet fields that involves $\alpha'$ is the $X(z)X(w)$ OPE, for dimensional reasons. Now, the correlator contractions $\Psi(z)\Psi(w)$ and $e^{ik\cdot X(z)}\partial (X/\alpha')(w)$ do not produce $\alpha'$ dependence. The contraction $\partial (X/\alpha')(z)\,\partial (X/\alpha')(w)$ does, but this is inverse dependence ($\sim 1/\alpha'$) and it also comes with $\partial_z g^{(1)}(z-w,\tau)$ as a result of the scalar Green's function; see section~\ref{sec:review} for the definition of the $g$ functions. We expect that all such contributions in the superstring correlator can be grouped and massaged into the following form:
\[
\label{eq:derinI}
\frac1{\alpha'}\,\text{KN}_n\, \partial_{z_i} (\cdots) \, = \text{discardable total derivatives} - \frac1{\alpha'}\, (\cdots) \,\partial_{z_i} \text{KN}_n\,,
\]
where the last term is independent of $\alpha'$ due to the form of the Koba-Nielsen factor. An explicit realisation of such a manipulation (though it started from a pure-spinor derivation) at one loop and 6 points was discussed in section 3.4.2 of \cite{Balli:2024wje}, motivated by the claims in \cite{Geyer:2024oeu}. An important point emphasised in \cite{Balli:2024wje} is that the function $(\cdots)$ must be single valued, so that the required cancellation occurs when integrating the total derivative over moduli space. We expect that this story extends to examples of multiple contractions at higher multiplicity. We also note that an expression like the left-hand side of \eqref{eq:derinI}, which includes $\partial_{z_i}g^{(1)}_{ij}$, does not fit the worldsheet basis we present, whereas the non-discardable term on the right-hand side does.

Let us make two more remarks. Firstly, the fact that $\mathcal{I}_n$ can be massaged to be independent of $\alpha'$ is reflected in the otherwise-unexpected relation of the $\alpha'\to\infty$ limit of the superstring integrand to its ambitwistor counterpart \cite{Mason:2013sva,Berkovits:2013xba,Huang:2016bdd,Casali:2016atr,Azevedo:2017yjy,Kalyanapuram:2021xow}, which describes the $\alpha'=0$ case of field theory. Secondly, we expect that this property of $\mathcal{I}_n$ extends to its higher-genus analogue. However, it is hard to anticipate what is the structure of the moduli-space integral at high enough genus, when chiral splitting is not possible.


\section{The IBP relation at 7 points}\label{app:ibp7pt}

In this appendix, we give the integration-by-parts (IBP) relations that exclude from the 7-point (that is, weight-3) worldsheet basis element any functions with closed cycles that appeared in the explicit pure-spinor computations of ref.~\cite{Mafra:2018pll}. This is the 7-point counterpart of the 6-point relation originally given in~\cite{Balli:2024wje} and shown also in~\eqref{eq:6ptibp}. Note that ref.~\cite{Mafra:2018pll} already listed IBP relations, but these used the total derivative of functions that are not single valued (i.e.~that are not monodromy invariant); this was later corrected in \cite{Balli:2024wje} at 6 points, and we will discuss 7 points here.

Refs.~\cite{Mafra:2018pll,Balli:2024wje} employed a convenient notion of covariant derivative with respect to a puncture, to denote a total derivative in moduli space:
\[
\nabla_i(f) \,\text{KN}_n = \partial_{z_i} (f\,\text{KN}_n)\,.
\]
In this language, the precise IBP relation we mentioned in \eqref{eq:6ptibp} is written as
\begin{equation}
    E_{1|2|3,4,5,6} = \frac{1}{s_{12}} \left( p_{1}^{\mu} p_2^{\nu} \, \reg\left\lbrack L_{\mu} L_{\nu} \right\rbrack + \sum_{i=3}^{n} p_{1}^{\mu} s_{2i} \, \reg\left\lbrack L_{\mu} V_{1|2,i} \right\rbrack \right) - \frac{1}{s_{12}} \nabla_2 (p_1\cdot L) \,,
\end{equation}
where $s_{ab} := \frac{1}{2} (p_a + p_b)^2 = p_a\cdot p_b$.

The worldsheet functions in open-string integrals with closed cycles that appeared in the 7-point pure-spinor calculation of ref.~\cite{Mafra:2018pll} are:
\begin{align}
    E_{1|23| 4,5,6,7}= & -s_{123} V^{(3)}_{1|2,3} + \frac{1}{2\alpha'} \left( g_{12}^{(1)}+g_{31}^{(1)} \right) \partial g_{23}^{(1)} + \frac{1}{2\alpha'} \partial g_{23}^{(2)} \,, \\
    E_{1|4| 23,5,6,7}= & \left[ \frac{1}{2\alpha'} \partial g_{14}^{(1)} - 2 s_{14} \eg{2}{14} + s_{14} (\egg{14})^2 \right] V_{1|2,3} -s_{24} V^{(3)}_{1|2,4} + s_{34} V^{(3)}_{1|3,4} \,, \\
    E_{1|2| 3,4,5,6,7}^{\mu}= & \left[ \frac{1}{2\alpha'} \partial g_{12}^{(1)} - 2 s_{12} \eg{2}{12} + s_{12} (\egg{12})^2  \right]\left(2\pi i \, \ell^{\mu}+\sum_{j \geq 3} k_j^{\mu} g_{1 j}^{(1)}\right)+\sum_{j \geq 3} k_j^{\mu} s_{2 j} V^{(3)}_{1|2,j} \\
    & + k_2^{\mu} \left[ \frac{1}{2\alpha'} \partial g_{12}^{(2)}+s_{12}\left(g_{12}^{(1)} g_{12}^{(2)}-3 g_{12}^{(3)}\right)\right] \,, \nonumber
\end{align}
where $s_{ijk\cdots} := \frac{1}{2} (p_i + p_j + p_k + \cdots)^2$, and $V^{(3)}$ is defined by
\begin{equation}
    V^{(3)}_{1|i,j} := \frac{1}{3!} \left( \egg{1i} + \egg{ij} - \egg{1j} \right)^{3} \bigg|_{(\egg{ab})^m \to m! \eg{m}{ab}} \,.
\end{equation}
For the closed-string integrals, the corresponding functions are obtained by the replacement $\alpha' \to \alpha'/4$.
The IBP relations needed to re-write them in terms of our basis elements are:
\begin{align}
    E_{1|4|23,5,6,7} =& \, \frac{1}{s_{14}} \reg\left\lbrack L^{(1)} L^{(4)} V_{1|2,3} \right\rbrack 
    - \frac{1}{s_{14}} \nabla_4 \reg \left\lbrack L^{(1)} V_{1|2,3} \right\rbrack \,, \\
    E^{\mu}_{1|2|3,4,5,6,7} =& \, \reg\left\lbrack \frac{1}{s_{12}} L^{\mu} L^{(1)} L^{(2)} - \frac{p^{\mu}_{2} }{2s_{12}^2} L^{(2)} (L^{(1)})^2 \right\rbrack + \nabla_2 \reg\left\lbrack \frac{p^{\mu}_{2}}{2s_{12}^2} (L^{(1)})^2 - \frac{1}{s_{1,2}} L^{\mu} L^{(1)} \right\rbrack \,, \nonumber \\
    E_{1|23|4,5,6,7} =& \, \frac{1}{2 s_{23}^2 (s_{12} + s_{13}) } \reg\left\lbrack L^{(2)} L^{(3)} ( s_{12} L^{(3)} - s_{13} L^{(2)} - 2 s_{23} (s_{12} + s_{13}) V_{1|2,3} ) \right\rbrack \\
    & + \left(\frac{1}{2 s_{23}^2 (s_{12} + s_{13}) } \nabla_2 \reg \left\lbrack s_{12} ( 2 s_{23} V_{1|2,3} - L^{(3)} ) L^{(3)} + 2 s_{23} L^{(1)} L^{(3)} \right\rbrack  - (2 \leftrightarrow 3)\right) \,, \nonumber
\end{align}
where
\begin{equation}
    L^{(a)} := \frac{1}{2\alpha'} \frac{\partial}{\partial z_a} \KN = 2\pi i\, \ell_{\mu} p_a^{\mu} + \sum_{i\neq a} s_{a,i} \egg{a,i} \,.
\end{equation}
Note that
\begin{equation}
    L^{(a)} = \left\{ 
    \begin{aligned}
        & p^{\mu}_{a} L_{\mu} \, && a = 1 \,, \\
        & p_a^{\mu} \Bigg\lbrack L_{\mu} + \sum_{i \neq 1,a} p_{i,\mu} V_{1|a,i} \Bigg\rbrack \,, && a \neq 1 \,,
    \end{aligned}
    \right.
\end{equation}
which makes it clear that the $L^{(a)}$'s still live in the space spanned by $L_{\mu}$ and the $V_{1|\cdots}$'s.


\section{Alternative regularisation involving $G_2(\tau)$}
\label{app:altreg}

We employ in the main body of the paper a worldsheet basis that, starting at 6 points, relies on a `regularisation' of closed cycles. This regularisation includes at higher points the appearance of the holomorphic Eisenstein series, $G_{2K}(\tau)$ for $K\geq2$, but not of its non-modular relative $G_2(\tau)$, defined in \eqref{eq:G2}. Here, we mention an alternative regularisation that puts $G_2(\tau)$ on equal footing with its modular counterparts. Naturally, $G_2(\tau)$ must still drop off the chiral integrand, as required by modularity, but it is interesting to see how this occurs in practice. Our standard regularisation, denoted by $\reg$ in the main text, will be replaced by the alternative regularisation $\reg'$.

Let us consider 6 points. The standard regularisation used in section~\ref{sec:6pt} acts as
\begin{equation}
    \begin{aligned}
        \reg\left\lbrack g_{i,j}^{(1)} g_{k,l}^{(1)} \right\rbrack &:= \egg{i,j}\egg{k,l} \,, \qquad \{ i,j \} \neq \{ k,l \} \,, \\
        \reg\left\lbrack \left( g_{1,i}^{(1)} \right)^2 \right\rbrack &:= \left( \egg{1,i} \right)^2 + \partial_1 g^{(1)}_{1,i} + G_2(\tau)= 2\eg{2}{1,i} \,,
    \end{aligned}
\end{equation}
whereas the alternative regularisation is
\begin{equation}
    \begin{aligned}
        \reg'\left\lbrack g_{i,j}^{(1)} g_{k,l}^{(1)} \right\rbrack &:= \egg{i,j}\egg{k,l} \,, \qquad \{ i,j \} \neq \{ k,l \} \,, \\
        \reg'\left\lbrack \left( g_{1,i}^{(1)} \right)^2 \right\rbrack &:= \left( \egg{1,i} \right)^2 + \partial_1 g^{(1)}_{1,i} = 2\eg{2}{1,i} - G_2(\tau) \,.
    \end{aligned}
\end{equation}
The later gives rise to $G_2(\tau)$.
The 6-point chiral integral in the standard regularisation is \eqref{eq:6ptreg}.
Exploiting the various relations between parts of the BCJ numerators, we can use the alternative regularisation to rewrite the chiral integrand as
\begin{equation}
\label{eq:I6altreg}
    \begin{aligned}
        \cI_{6} = &\; N_{6}^{\mu\nu} \, \reg'\left\lbrack L_{(\mu} L_{\nu)} \right\rbrack - 
        \sum_{2 \leqslant i<j \leqslant 6} N_{6}^{\mu}([i,j]) \, \reg' \left\lbrack L_{\mu} \, V_{1|i,j} \right\rbrack \\
        & + \sum_{ \substack{2 \leq i < j , k \leq 6 \\  j \neq k } } N_6([[i,j],k])\, V_{1|i,j,k} + \sum_{ \substack{ 2 \leqslant i<j \leqslant 6 \\ 2 \leqslant i<k<l \leqslant 6 \\ j \neq k,l } } N_6([i,j],[k,l]) \, V_{1|i,j} V_{1|k,l} \\
        & + \frac{(2\pi i)^2}{5!} \sum_{\rho\in S_5} N\big(1 \rho(2)\cdots \rho(6); -\frac{p_1}{2}\big)\; \frac{G_2(\tau)}{2\zeta(2)}  \,. 
    \end{aligned}
\end{equation}
Recollecting the total coefficient of $G_2(\tau)$ --- which comes not only from the last line but also from the use of $\reg'$ --- and requiring it to vanish leads exactly to the identity \eqref{eq:6ptidG2}, and reduces the expression above to \eqref{eq:6ptreg}. 

We see that inclusion of $G_2(\tau)$ at 6 points is analogous to the inclusion of $G_4(\tau)$ at 8 points in \eqref{eq:8ptreg}, and also of $G_{2K}(\tau)$ at $4+2K$ points. So while this alternative regularisation is obtuse, it does show that $G_2(\tau)$ fits in nicely before we impose actual modularity. At arbitrary multiplicity, the prescription for $\reg'$ is exactly as the one for $\reg$ presented in section~\ref{subsec:reg}, except that we do not perform the final step \eqref{eq:finalreg} where we eliminate the remaining appearance of $G_2(\tau)$.


\section{Sketch of proof of worldsheet basis}\label{app:proof}

In this appendix, we will sketch a brief proof of the ansatz for the superstring chiral integrand presented in section~\ref{sec:npt}, by assuming that all the closed cycles can be ruled out. Since the classical modular forms denoted as $\gamma_{2K}(\tau)$ in \eqref{eq:basis} are monodromy invariant \textit{per se}, we will only talk about the accompanying monodromy-invariant functions of loop momentum and punctures, that is, the objects $\reg[L_{\mu_1}\cdots L_{\mu_{w_L}}V\cdots V]$. 

We start by fixing the $\ell$-dependent part of the chiral integrand. We will study it recursively by multiplicity. Suppose the $(n-1)$-point integrand is given. According to the discussion in our previous work~\cite[section 8.1]{Geyer:2024oeu}, the $\ell$-dependent part of the $n$-point integrand can be fully fixed from the $(n-1)$-point integrand, as they are structurally the same, with the former one containing one more factor of $2\pi i\,\ell$. One can then reorganise the $\ell$-dependent part into a regularised polynomial of $L_{\mu}$ (and $V$'s) minus $\ell$-independent terms. Therefore, we can separate the $n$-point integrand into two parts:
\begin{equation}
    \cI_n=\reg\big\lbrack  L_{\mu}\big(\textrm{($n-1$)-point structure}\big)^\mu \big\rbrack + (\ell\text{-independent monodromy-invariant function}) \,. 
\end{equation}
So now only the second part above is unknown. We will show recursively (in multiplicity) that this piece can be expanded into the basis~\eqref{eq:basis} with $w_L=K=0$, namely products of the form $\prod V_{1|\cdots}$\,. Unlike in section~\ref{subsec:basis}, here we should start from an integrand of arbitrary fixed weight.\footnote{It is necessary to consider this larger space of arbitrary weight. The reason will become clear in the proof below.} The corresponding basis is still composed of~\eqref{eq:basis} with $w_L=K=0$, but now without any restriction on $v$.\footnote{Therefore, such basis will also work for the chiral integrand of non-maximally supersymmetric string theories.} We will establish this claim in the proof below.

By the Fay identity \eqref{eq:faygeneral}, setting $j=n$, we can expand an $\ell$-independent $n$-point integrand with weight $w$, denoted by $\cI_{n,0}^{(w)}$, into monomials that include at most one instance of $\eg{\bullet}{\bullet,n}$:
\begin{equation}\label{eq:ansatzsingle}
    \cI_{n,0}^{(w)} = f^{(w)}(z_1, \cdots, z_{n-1}) + \sum_{i=1}^{n-1} \sum_{m=1}^{w} \eg{m}{i,n} f^{(w-m)}_{i}(z_1, \cdots, z_{n-1}) \,,
\end{equation}
where every function $f$ is free of closed cycles. Recall that the whole function should be monodromy invariant. Considering the monodromy transformation $z_n \to z_n+\tau$ of the ansatz above, we are led to:
\begin{equation}
    0 = \sum_{i=1}^{n-1} \sum_{m=1}^{w} \left( \sum_{k=0}^{m-1} \frac{(-2\pi i)^{m-k}}{(m-k)!} \eg{k}{i,n} \right) f^{(w-m)}_{i}(z_1, \cdots, z_{n-1}) \,.
\end{equation}
Solving the equation for arbitrary $z_n$, we obtain
\begin{equation}
    \begin{aligned}
        & f^{(w-m)}_{i}(z_1, \cdots, z_{n-1}) = 0 \,, \quad \forall\, m > 1\,, \\
        & \sum_{i=1}^{n-1} f^{(w-1)}_{i}(z_1, \cdots, z_{n-1}) = 0 \,.
    \end{aligned}
\end{equation}
Inserting the solution above back into the ansatz~\eqref{eq:ansatzsingle}, we get
\begin{equation}\label{eq:proofansatzreg}
    \begin{aligned}
        \cI_{n,0}^{(w)} &= f^{(w)}(z_1, \cdots, z_{n-1}) + \sum_{i=2}^{n-1} \big(\eg{1}{i,n} - \eg{1}{1,n}\big) f^{(w-1)}_{i}(z_1, \cdots, z_{n-1})  \,. \\
        &= \tilde{f}^{(w)}(z_1, \cdots, z_{n-1}) + \sum_{i=2}^{n-1} \big(\eg{1}{i,n} - \eg{1}{1,n} - \eg{1}{1,i} + \hat{D}_i\big) f^{(w-1)}_{i}(z_1, \cdots, z_{n-1})\bigg|_{G_{2K} \mapsto 0}  \,,
    \end{aligned}
\end{equation}
with
\begin{equation}
    \tilde{f}^{(w)}(z_1, \cdots, z_{n-1}) := f^{(w)}(z_1, \cdots, z_{n-1}) + \sum_{i=2}^{n-1} \left( \eg{1}{1,i} - \hat{D}_i \right) f^{(w-1)}_{i}(z_1, \cdots, z_{n-1}) \bigg|_{G_{2K} \mapsto 0} \,, 
    \label{eq:tildef}
\end{equation}
and $\hat{D}_i$ defined in~\eqref{eq:fulldhat}. The rewriting in the second line of~\eqref{eq:proofansatzreg} serves two purposes. 
Firstly, the monodromy transformation becomes easier to handle: for any $a$ such that $ 1 \leqslant a \leqslant n-1$, we have
\begin{equation}
    \begin{aligned}
        \cI_{n,0}^{(w)} \Big|_{z_a \to z_a + \tau} =& \left\lbrack \tilde{f}^{(w)}(z_1, \cdots, z_{n-1})  \Big|_{z_a \to z_a + \tau} \right\rbrack  \\
        & + \sum_{i=2}^{n-1} \big(\eg{1}{i,n} - \eg{1}{1,n} - \eg{1}{1,i} + \hat{D}_i\big) \left\lbrack f^{(w-1)}_{i}(z_1, \cdots, z_{n-1}) \Big|_{ z_a \to z_a + \tau } \right\rbrack \bigg|_{ G_{2K} \mapsto 0 } \,.
    \end{aligned}
\end{equation}
Monodromy invariance then implies that $\tilde{f}^{(w)}$ and all the $f_i^{(w-1)}$ must be monodromy invariant individually. Secondly, adding $\hat{D}_i$ along every instance of $\egg{1,i}$ in \eqref{eq:tildef} ensures that $\tilde{f}^{(w)}$ is free of closed cycles. 

Therefore, we conclude that $\tilde{f}^{(w)}$ and each $f_{i}^{(w-1)}$ is an $(n-1)$-point monodromy invariant function with weights $w$ and $(w-1)$, respectively. Eq.~\eqref{eq:proofansatzreg} thus provides a recursion relation expressing an $n$-point integrand of weight $w$ in terms of $(n-1)$-point integrands of weight $w$ and $(w-1)$. We assume that the $(n-1)$-point integrand is known to be spanned by the basis~\eqref{eq:basis} with $w_L = K = 0$, i.e.~by products $ \prod_{r=1}^v V_{1|i_1^{(r)}i_2^{(r)}\cdots i_{s_r}^{(r)}}$ of the appropriate weight. The $n$-point integrand is then spanned by a basis of the form:
\begin{equation}\label{eq:nptbasis}
    \begin{aligned}
        & \prod_{r=1}^v V_{1|i_1^{(r)}i_2^{(r)}\cdots i_{s_r}^{(r)}} \,, \\
        & \big(\eg{1}{i,n} - \eg{1}{1,n} - \eg{1}{1,i} + \hat{D}_i\big) \prod_{r=1}^v V_{1|i_1^{(r)}i_2^{(r)}\cdots i_{s_r}^{(r)}} \bigg|_{G_{2K} \mapsto 0} \,, \\
    \end{aligned}
\end{equation}
where the $V$'s here only contain $(n-1)$ punctures.
The first line above is automatically in the form of products. 
For the second line, if the puncture $i$ does not belong to any set $\{ i^{(r)}_{\bullet} \}$ in the product of $V$'s, then we simply have
\begin{equation}
    \big(\eg{1}{i,n} - \eg{1}{1,n} - \eg{1}{1,i} + \hat{D}_i\big) \prod_{r=1}^v V_{1|i_1^{(r)}i_2^{(r)}\cdots i_{s_r}^{(r)}} \bigg|_{G_{2K} \mapsto 0} = - V_{1|i,n} \prod_{r=1}^v V_{1|i_1^{(r)}i_2^{(r)}\cdots i_{s_r}^{(r)}} \,.
\end{equation}
On the other hand, if the puncture $i$ belongs to one set $\{i^{(r)}_{\bullet}\}$, e.g.~we take $i = i^{(m)}_{k}$, we get
\begin{equation}
    \begin{aligned}
        & \Big(\eg{1}{i^{(m)}_{k},n} - \eg{1}{1,n} - \eg{1}{1,i^{(m)}_{k}} + \hat{D}_{i^{(m)}_{k}}\Big) \prod_{r=1}^v V_{1|i_1^{(r)}i_2^{(r)}\cdots i_{s_r}^{(r)}} \bigg|_{G_{2K} \mapsto 0} \\
        =& \sum_{p = k}^{s_m} V_{1|i_1^{(m)} \cdots i_{p}^{(m)} n\, i_{p+1}^{(m)} \cdots i_{s_m}^{(m)}} \prod_{r \neq m} V_{1|i_1^{(r)}i_2^{(r)}\cdots i_{s_r}^{(r)}} \,.
    \end{aligned}
\end{equation}
So the second line of~\eqref{eq:nptbasis} also falls into the form of products of $V$'s. Therefore the $n$-point $\ell$-independent integrand can be spanned by the basis of the form~\eqref{eq:basis} with $w_L = K = 0$, thereby completing the proof.

\bibliography{twistor-bib}
\bibliographystyle{JHEP}

\end{document}